\documentclass[fleqn,usenatbib]{mnras}

\usepackage{newtxtext,newtxmath}

\usepackage[T1]{fontenc}

\DeclareRobustCommand{\VAN}[3]{#2}
\let\VANthebibliography\thebibliography
\def\thebibliography{\DeclareRobustCommand{\VAN}[3]{##3}\VANthebibliography}

\usepackage[dvipdfmx]{graphicx}	
\usepackage{amsmath}	
\usepackage{hyperref}
\usepackage{bm}
\usepackage{comment}
\usepackage{here}
\usepackage[dvipsnames]{xcolor}
\usepackage{multicol}
\usepackage{lscape}

\newcommand{\hMpc}{h^{-1} \, \mathrm{Mpc}}

\title[Structure of troughs with Subaru HSC]{Line-of-sight structure of troughs identified
in Subaru Hyper Suprime-Cam Year 3 weak lensing mass maps}

\author[T. Shimasue et al.]{
Takumi Shimasue$^{1,2,3}$\thanks{E-mail: shimasue-takumi@resceu.s.u-tokyo.ac.jp},
Ken Osato$^{4,3,5}$,
Masamune Oguri$^{4,3,5}$,
Rhythm Shimakawa$^{6}$,
\newauthor
Atsushi J. Nishizawa$^{7,8}$
\\
$^{1}$Department of Physics, Graduate School of Science, The University of Tokyo,
7-3-1 Hongo, Bunkyo, Tokyo 113-0033, Japan\\
$^{2}$Research Center for the Early Universe, School of Science, The University of Tokyo,
7-3-1 Hongo, Bunkyo, Tokyo 113-0033, Japan\\
$^{3}$Department of Physics, Graduate School of Science, Chiba University,
1-33 Yayoicho, Inage, Chiba 263-8522, Japan\\
$^{4}$Center for Frontier Science, Chiba University,
1-33 Yayoicho, Inage, Chiba 263-8522, Japan\\
$^{5}$Kavli Institute for the Physics and Mathematics of the Universe,
The University of Tokyo Institutes for Advanced Study,\\
5-1-5 Kashiwanoha, Kashiwa, Chiba 277-8583, Japan\\
$^{6}$Waseda Institute for Advanced Study (WIAS), Waseda University,
1-21-1 Nishi-Waseda, Shinjuku, Tokyo 169-0051, Japan\\
$^{7}$DX center, Gifu Shotoku Gakuen University, 1-1 Takakuwanishi, Yanaizucho, Gifu 501-6194, Japan\\
$^{8}$Kobayashi Maskawa Institute/Institute for Advanced Research, Nagoya University, Furocho, Chikusa, Nagoya, Aichi 464-8602, Japan
}

\date{Accepted XXX. Received YYY; in original form ZZZ}

\pubyear{2023}

\begin{document}
\label{firstpage}
\pagerange{\pageref{firstpage}--\pageref{lastpage}}
\maketitle

\begin{abstract}
We perform the weak lensing mass mapping analysis
to identify \textit{troughs}, which are defined as local minima in the mass map.
Since weak lensing probes the projected matter distribution along the line-of-sight,
these troughs can be produced by single or multiple voids projected along the line-of-sight.
To scrutinise the origins of the weak lensing troughs,
we systematically investigate the line-of-sight structure of troughs selected from the latest Subaru Hyper Suprime-Cam (HSC) Year 3 weak lensing data covering $433.48 \, \mathrm{deg}^2$.
From a curved sky mass map constructed with the HSC data,
we identify 15 troughs with the signal-to-noise ratio higher than $5.7$ and
address their line-of-sight density structure 
utilizing redshift distributions of two galaxy samples, 
photometric luminous red galaxies observed by HSC and spectroscopic galaxies
detected by Baryon Oscillation Spectroscopic Survey.
While most weak lensing signals due to the troughs are explained by multiple voids aligned along the line-of-sight, we find that two of the 15 troughs potentially originate from single voids at redshift $\sim 0.3$. The single void interpretation appears to be consistent with our three-dimensional mass mapping analysis. We argue that single voids can indeed reproduce observed weak lensing signals at the troughs if these voids are not spherical but are highly elongated along the line-of-sight direction.
\end{abstract}




\section{Introduction}
\label{intro}
The shapes and sizes of distant galaxies are distorted due to the gravitational potential sourced by the intervening matter distribution.
This phenomenon is called weak gravitational lensing.
In particular, the coherent shape distortions induced by the large-scale structure of the Universe
are referred to as \textit{cosmic shear} \citep[for reviews, see][]{Bartelmann2001,Kilbinger2015,Mandelbaum2018}.
Since cosmic shear measurements and analyses do not rely on the uncertain relationship
between galaxies and matter as adopted in galaxy clustering analysis,
the matter distribution can be probed in an unbiased manner.
Thus, cosmic shear is recognized as a powerful and unique approach to studying the nature of dark matter and dark energy in modern cosmology.

The standard approach to extracting cosmological information from weak lensing measurements
is the correlation analysis of galaxy shapes.
The widely used statistics are two-point statistics,
i.e., the correlation function and power spectrum,
which are sensitive to the dark matter density and the fluctuation amplitude of matter inhomogeneities
and contain complete information for the Gaussian random field \citep{Schneider2002}.
While the matter distribution in the very early Universe is close to Gaussian,
the late-time matter density field becomes highly non-Gaussian due to the non-linear gravitational growth of the matter distribution.
To capture the non-Gaussian information,
statistics beyond two-point correlations are required.
Among them, peak statistics of weak lensing mass maps
\citep{Jain2000,Hamana2004,Yang2011,Liu2016,Fluri2018},
i.e., the number density of peaks as a function of peak height, is thought to be one of the most promising statistics since weak lensing peaks are easily identified, and high peaks have clear correspondence
with the density peaks of the matter distribution \citep[see e.g.,][]{Miyazaki2018b,Oguri2021}.
Indeed, the peak statistics have been employed
to constrain cosmological parameters in weak lensing measurements \citep{Liu2015,Kacprzak2016,Shan2018,Harnois-Deraps2021}.

Because of their potential association with \textit{voids} \citep[for a review, see][]{vandeWeygaert2011}, the low-density region in the Universe, \textit{troughs},
which are defined as local minima in weak lensing mass maps,
are also expected to convey cosmological information \citep{Jain2000,Miyazaki2002}.
The first identifications of the void have been reported in \citet{Gregory1978,Joeveer1978}
followed by detections with larger spectroscopic surveys \citep{Kirshner1981,deLapparent1986}.
As a recent study, \citet{Douglass2023} present the void catalogue constructed from
Sloan Digital Sky Survey (SDSS) Main Sample in Data Release 7,
in which more than 1,000 voids with radii larger than $10 \, \hMpc$ are identified.
Such a large sample of voids enables a robust measurement of
the void statistics, which offers a new avenue to study modified gravity theories \citep[e.g.,][]{Cai2015,Nadathur2016}.
The structure and abundance of voids are less sensitive to baryonic physics due to the scarcity of gas and the fact that the formation and evolution are purely driven by gravity.
The clustering analysis of voids is employed to constrain the geometry of the Universe
thorugh the Alcock--Paczynski test
\citep{Lavaux2012,Hamaus2016,Mao2017,Nadathur2019,Hamaus2020} and
to measure the baryon acoustic oscillation scale \citep{Kitaura2016,Liang2016,Nadathur2019,Zhao2020,Zhao2022}.
Accordingly, the same feature is expected to hold for weak lensing trough statistics,
and the trough statistics contain information complementary to peak statistics
\citep{Gruen2016,Barreira2017,Coulton2020,Davies2021,Osato2021}.

The voids have attracted a lot of attention due to their potential association with the \textit{cold spot} of cosmic microwave background (CMB),
which is a large low-temperature region
located at $(l, b) \simeq (209\degr, -57\degr)$ in the Galactic coordinate.
The cold spot is first reported by Wilkinson Microwave Anisotropy Probe \citep{Bennett2013}
and further confirmed by \textit{Planck} \citep{Planck2013XXIII}.
One of the plausible explanations for the origin of the cold spot
is a \textit{supervoid}, which is a large-scale ($\gtrsim 100 \, \hMpc$) underdense region
and possibly consists of multiple voids \citep{Inoue&Silk2006}.
The decaying gravitational potential of the supervoid can generate the decrement of
the CMB temperature through the integrated Sachs--Wolfe effect.
Indeed, at the position of the cold spot, an extended void region ($\simeq 200 \, \hMpc$) called 
the Eridanus supervoid is identified at the redshift $z \simeq 0.2$  
\citep{Szapudi2015,Kovacs2022}.
However, the shallow underdensity of the Eridanus supervoid with the density contrast of $\delta \simeq - 0.2$
can account for only $10\text{--}20$ per cent of the decrement signal
assuming the $\Lambda$ cold dark matter cosmological model.
A further extensive study of the supervoid region is required to fully confirm the hypothesis that the origin of the cold spot is the supervoid.

The straightforward method to find underdense regions in the Universe is
based on the galaxy number density field, which is the biased tracer of matter,
measured from spectroscopic surveys.
However, voids are regions with fewer or no galaxies by nature, and thus,
the identification of voids from galaxy catalogues entails large statistical uncertainty.
Furthermore, the different void-finding algorithms
\citep{El-Ad1997,Hoyle2002,Neyrinck2008,Sutter2015,Nadathur2019}
lead to different void populations, resulting in large systematic noise. The comparison of the observed void catalogue with simulations also depends on how galaxies are populated in the simulations, and therefore is subject to the baryon physics uncertainty.

The void search in weak lensing mass maps has a potential to identify underdense regions that galaxy surveys cannot detect and settle the debate about the origin of the cold spot. However, simple analytic estimates indicate that single voids cannot produce significant weak lensing signals unless their size is extremely large, $\gtrsim 100 \, \hMpc$ \citep{Amendola1999}. 
Therefore, it is likely that the detection of a single large underdense region is difficult with weak lensing.
This is why previous work of weak lensing by voids has primarily focused on the stacked weak lensing analysis of voids identified from galaxy distributions \citep{Higuchi2013,Krause2013,Melchior2014,Clampitt2015,Sanchez2017,Fang2019,Vielzeuf2021}. Troughs in weak lensing mass maps tend to be associated with multiple underdense regions along the line-of-sight \citep[see, e.g.,][]{Chang2018}, although possible weak lensing troughs associated with single large voids have also been identified \citep[see, e.g.,][]{Jeffrey2021,Shimakawa2021}. In either case, studies of the line-of-sight structure of weak lensing troughs have been limited to those in a few curious cases, and there have not been any systematic studies of the line-of-sight structures of weak lensing troughs.

In this paper, we employ the latest Subaru Hyper Suprime-Cam (HSC) Year 3 (Y3) weak lensing shape catalogue
\citep{Li2022} to systematically study line-of-sight structures of the most significant troughs identified in weak lensing mass maps. The weak lensing shape catalogue spans $433.48 \, \mathrm{deg}^2$ with the mean source galaxy number density $22.9 \, \mathrm{arcmin}^{-2}$.
The high number density from the deep imaging data enables us to probe the density field out to higher redshifts. It offers an ideal tool to map the cosmic web structure with high statistical significance.
We study the line-of-sight structures of the most significant troughs employing two galaxy catalogues;
one is the photometric luminous red galaxy (LRG) catalogue selected with CAMIRA algorithm \citep{Oguri2014,Oguri2018a,Oguri2018} and the other is the LOWZ and CMASS spectroscopic galaxy samples from Baryon Oscillation Spectroscopic Survey of SDSS Data Release 12 \citep{Reid2016}.
Furthermore, we perform the three-dimensional weak lensing mass mapping \citep{Simon2009,Oguri2018}
to probe the large-scale density field around the identified troughs.

This paper is organized as follows.
In Section~\ref{sec:mass_mapping},
we briefly overview the basics of weak lensing and the mass mapping analysis.
In Section~\ref{sec:catalogues},
we present the weak lensing shape catalogue from HSC Y3 data and galaxy catalogues
to identify the weak lensing troughs and investigate the line-of-sight structures at trough positions.
In Section~\ref{sec:results}, we present the results on identifying weak lensing troughs
and line-of-sight galaxy number densities at the trough positions.
We discuss our results in Section~\ref{sec:discussions} and give conclusions in Section~\ref{sec:conclusions}.
Throughout this paper, we adopt a flat $\Lambda$ cold dark matter cosmology with
the matter density $\Omega_\mathrm{m} = 0.3$,
the baryon density $\Omega_\mathrm{b} = 0.05$,
the Hubble constant $H_0 = 100h = 70 \, \mathrm{km} \mathrm{s}^{-1} \, \mathrm{Mpc}^{-1}$,
the tilt of the scalar perturbation $n_\mathrm{s} = 0.96$, and the present amplitude of the matter fluctuation at the scale of $8\hMpc$ $\sigma_8 = 0.81$.

\section{Weak lensing mass mapping with Subaru HSC Y3 data}
\label{sec:mass_mapping}
This Section overviews the basics of weak lensing mass mapping analysis
and the HSC Y3 shear catalogue. Throughout the paper, weak lensing mass maps are constructed in a curved sky without adopting the flat sky approximation, in contrast to the previous weak lensing mass map analyses using the HSC survey data \citep{Oguri2018,Oguri2021,Miyazaki2018b} for which the flat sky approximation has been adopted.

\subsection{Two-dimensional mass mapping}
To formulate weak lensing, we begin by defining the lensing potential
for a single source located at the comoving distance of $\chi_\mathrm{s}$:
\begin{equation}
    \psi (\chi_\mathrm{s}, \bm{\theta}) = \frac{2}{c^2} \int_{0}^{\chi_\mathrm{s}} \mathrm{d}\chi
    \frac{f_K(\chi_\mathrm{s}-\chi)}{f_{K}(\chi_\mathrm{s})f_K(\chi)}\Phi (\chi, \bm{\theta}),
\end{equation}
where $\bm{\theta}=(\theta, \phi)$ specifies the position on the sky,
$c$ is the speed of light,
$\Phi$ is the gravitational potential,
and $f_K (\chi)$ is the comoving angular diameter distance
with the curvature $K$:
\begin{equation}
f_K (\chi) =
  \begin{cases}
    \sin (\sqrt{K} \chi) / \sqrt{K} & (K>0) , \\
    \chi & (K=0) , \\
    \sinh (\sqrt{-K} \chi) / \sqrt{-K} & (K<0) ,
  \end{cases}
\end{equation}
although we note that the flat Universe ($K = 0$)
is assumed throughout the paper.
The gravitational potential can be derived from the Poisson equation:
\begin{equation}
    \nabla_x^2 \Phi (\chi, \bm{\theta})
    = \frac{3 \Omega_\mathrm{m} H_0^2}{2 a} \delta (\chi, \bm{\theta}),
\end{equation}
where $\nabla_x$ is the differential operator in the comoving coordinate,
$a$ is the scale factor, and $\delta (\chi, \bm{\theta})$ is the density contrast.
The convergence $\kappa$ and shear $\gamma$ fields,
which correspond to the isotropic and anisotropic deformations of images, respectively,
are introduced in the curved sky \citep{Heavens2003,Castro2005,Chang2018,Jeffrey2021}:
\begin{equation}
\label{eq:kappa_gamma}
    \kappa = \frac{1}{4} (\eth \bar{\eth} + \bar{\eth} \eth) \psi ,\
    \gamma = \frac{1}{2} \eth \bar{\eth} \psi, 
\end{equation}
where $\eth$ and $\bar{\eth}$ are the diffrential raising and lowering operators in spin-$s$ spherical harmonics ${}_s Y_{\ell m}$.
The convergence field $\kappa (\bm{\theta}, \chi_\mathrm{s})$ with the single source
at the comoving distance $\chi_\mathrm{s}$ can be computed as
\begin{multline}
\kappa (\bm{\theta}, \chi_\mathrm{s}) = \\
\frac{3 H_0^2 \Omega_\mathrm{m}}{2 c^2}
\int_0^{\chi_\mathrm{s}} \!\! \frac{\mathrm{d} \chi}{a(\chi)}
\frac{f_K (\chi_\mathrm{s} - \chi) f_K (\chi)}{f_K (\chi_\mathrm{s})} 
\delta (f_K (\chi) \bm{\theta}, \chi).
\end{multline}
Hence, the convergence field is the projected density contrast with the distance kernel, referred to as the \textit{mass map}.
In real measurements, we use multiple source galaxies with different redshifts and
therefore the observed convergence field $\kappa (\bm{\theta})$ should be
weighted with the redshift distribution:
\begin{equation}
\kappa (\bm{\theta}) = \int_0^{\chi_\mathrm{H}} \!\! \mathrm{d} \chi_\mathrm{s} \, p(\chi_\mathrm{s})
\kappa (\bm{\theta}, \chi_\mathrm{s}) ,
\end{equation}
where $\chi_\mathrm{H}$ is the comoving distance to the horizon and $p(\chi_\mathrm{s})$ is the redshift distribution of source galaxies normalised as
$\int_0^{\chi_\mathrm{H}} \mathrm{d}\chi_\mathrm{s} \, p(\chi_\mathrm{s}) = 1$.
The expression can be recast as
\begin{equation}
\kappa (\bm{\theta}) = \frac{3 H_0^2 \Omega_\mathrm{m}}{2 c^2}
\int_0^{\chi_\mathrm{H}} \!\! \frac{\mathrm{d} \chi}{a(\chi)} q(\chi) f_K (\chi)
\delta (f_K (\chi) \bm{\theta}, \chi) ,
\end{equation}
where the lens efficiency $q (\chi)$ is defined as
\begin{equation}
q(\chi) = \int_\chi^{\chi_\mathrm{H}} \!\! \mathrm{d} \chi_\mathrm{s} \,
p(\chi_\mathrm{s}) \frac{f_K (\chi_\mathrm{s} - \chi)}{f_K (\chi_\mathrm{s})} .
\end{equation}
The shear and convergence fields are not independent but
related in the harmonic space.
The harmonic expansions of
the lensing potential, convergence and shear fields are given as
\begin{align}
    \psi &= \sum_{\ell m}
    \tilde{\psi}_{\ell m} \, Y_{\ell m} (\theta, \phi), \\
    \kappa &= \sum_{\ell m}
    \tilde{\kappa}_{\ell m} \, Y_{\ell m} (\theta, \phi), \\
    \gamma &= \sum_{\ell m} \tilde{\gamma}_{\ell m} \, {}_2 Y_{\ell m} (\theta, \phi),
\end{align}
where $\tilde{\psi}$, $\tilde{\kappa}$, and $\tilde{\gamma}$ are
the lensing potential, convergence, and shear in harmonic space, respectively, $Y_{\ell m}$ is the spherical harmonics,
and ${}_2 Y_{\ell m}$ is the spin-2 spherical harmonics.
By plugging the harmonic expansions into Eq.~\eqref{eq:kappa_gamma},
the convergence and shear fields and the lensing potential
are related linearly:
\begin{align}
    \tilde{\kappa}_{\ell m} &= -\frac{1}{2} \ell (\ell + 1) \tilde{\psi}_{\ell m}, \\
    \tilde{\gamma}_{\ell m} &= \frac{1}{2}
    \sqrt{(\ell -1) \ell (\ell +1) (\ell + 2)} \tilde{\psi}_{\ell m} \\
    &= -\sqrt{\frac{(\ell -1) (\ell + 2)}{\ell (\ell + 1)}} \tilde{\kappa}_{\ell m}
    \label{eq:kappa_sht}.
\end{align}
Thus, the convergence field, or equivalently, the mass map, can be obtained
once the shear field is estimated from the shape catalogue.
The real and imaginary parts of convergence and shear field in harmonic space
are referred to as E-mode and B-mode, respectively:
\begin{align}
\tilde{\kappa}_{\ell m} &= \tilde{\kappa}_{E, \ell m} + i \tilde{\kappa}_{B, \ell m}, \\
\tilde{\gamma}_{\ell m} &= \tilde{\gamma}_{E, \ell m} + i \tilde{\gamma}_{B, \ell m} .
\end{align}
Since the scalar potential $\psi$ induces weak lensing effect,
the E-mode convergence $\kappa_E$ prevails over the B-mode convegence $\kappa_B$.
However, the systematic effect, such as incomplete point spread function correction
results in non-zero B-mode convergence.
Thus, the B-mode convergence should be zero within statistical uncertainty
and can be used as a null test.

In the practical analysis of weak lensing,
we estimate the shear field $\hat{\gamma}_{\alpha} (\bm{\theta})$ as
\begin{equation}
    \label{eq:shear_from_shapes}
    \hat{\gamma}_{\alpha} (\bm{\theta}) =
    \frac{\sum_i w_i (\gamma_\alpha (\bm{\theta}_i) - c_{\alpha,i}) W(\bm{\theta}; \bm{\theta}_i)}{\sum_i w_i(1+m_i) W(\bm{\theta}; \bm{\theta}_i)} ,
\end{equation}
where the sum runs over all source galaxies,
$\gamma_\alpha (\bm{\theta}_i)$ is the local shear field at the position of the $i$-th galaxy,
$w_i$ is the lensing weight \citep{Mandelbaum2018}, $c_{\alpha, i}$ is the additive bias, $m_i$ is the multiplicative bias (see Section~\ref{sec:HSC_shape_catalogue}),
$W$ is the Gaussian smoothing kernel.
The local shear field is estimated as
\begin{align}
    \gamma_\alpha (\bm{\theta}_i) = \frac{e_\alpha (\bm{\theta}_i)}{ 2 \mathcal{R}},
\end{align}
where $e_\alpha (\bm{\theta}_i)$ is the galaxy shape ellipticity and
$\mathcal{R}$ is the shear responsivity given as
\begin{equation}
    \mathcal{R} = 1 - \frac{\sum_i w_i e_{\mathrm{rms}, i}^2}{\sum_i w_i},
\end{equation}
where $e_{\mathrm{rms}, i}$ is the intrinsic shape dispersion.
The smoothed shear field is estimated from the shape catalogue (Eq.~\ref{eq:shear_from_shapes})
and pixellated based on the \texttt{HEALPix} \citep{Gorski2005} pixellization.
Spherical harmonic coefficients of the estimated shear
are then calculated with the \texttt{map2alm\_spin} routine of \texttt{healpy} \citep{Zonca2019}.
Finally, the convergence field can be computed from the smoothed shear field in harmonic space through the relation (Eq.~\ref{eq:kappa_sht}).
We mask pixels with few galaxies because the shear estimate in such pixels suffers
from a large statistical uncertainty.
In practice, we construct the source galaxy number density-weighted with lensing weights
and the pixels with less than $20$ per cent of the mean of the number density map
are masked.

The size of the Gaussian smoothing kernel $W$ should be carefully considered. In principle, this smoothing kernel size sets the minimum size of troughs or voids we can find from mass maps. On the one hand, a large kernel size is desired to find large voids or supervoids. Still, on the other hand, the effects of residual systematics arising from imperfect shape measurements on weak lensing mass maps become more critical when the smoothing size is larger \citep{Oguri2018,Li2022}. The large smoothing size also significantly reduces the effective search area because many troughs touch the edge of the survey footprint and, therefore, will be discarded.
In addition, an advantage of the HSC survey is its high number density of source galaxies, which enables us to construct mass maps down to smaller scales accurately. In this paper, we choose the Gaussian smoothing kernel with the full-width half maximum (FWHM) scale of $40 \, \mathrm{arcmin}$ ($\simeq 11 \, \hMpc$ at $z = 0.2$) as a compromise between the size of troughs and voids and possible systematic errors in weak lensing mass maps. Note that the analysis is repeated with the smoothing scale of $20 \, \mathrm{arcmin}$ and $60 \, \mathrm{arcmin}$ to confirm that our results are insensitive to the choice of the smoothing size.  In our analysis, we adopt the \texttt{Healpix} pixelization with $N_\mathrm{side} = 512$, which is sufficient to resolve the smoothing kernel.

To estimate the statistical significance of weak lensing mass maps,
we construct the signal-to-noise ratio (S/N) map defined as the convergence map divided by the standard deviation of the random noise maps \citep{Oguri2018}. First, we create random mass maps, which are mass maps computed from randomly rotated galaxy shapes, and repeat generating random noise maps $100$ times by changing the random seed. The random rotation erases the cosmological signal and thus, random noise maps reflect the statistical uncertainties of mass mapping. Then, we measure the standard deviation of the $100$ random mass maps, which corresponds to the statistical noise at the pixel level. Finally, we obtain the S/N maps by dividing the convergence map with the noise standard deviation map.
We define troughs as pixels in the S/N map that are lower than all neighbour pixels.
We note that the convergence at troughs is negative in most cases,
and correspondingly, S/N values of troughs are negative.

\begin{figure}
    \includegraphics [width=\columnwidth]{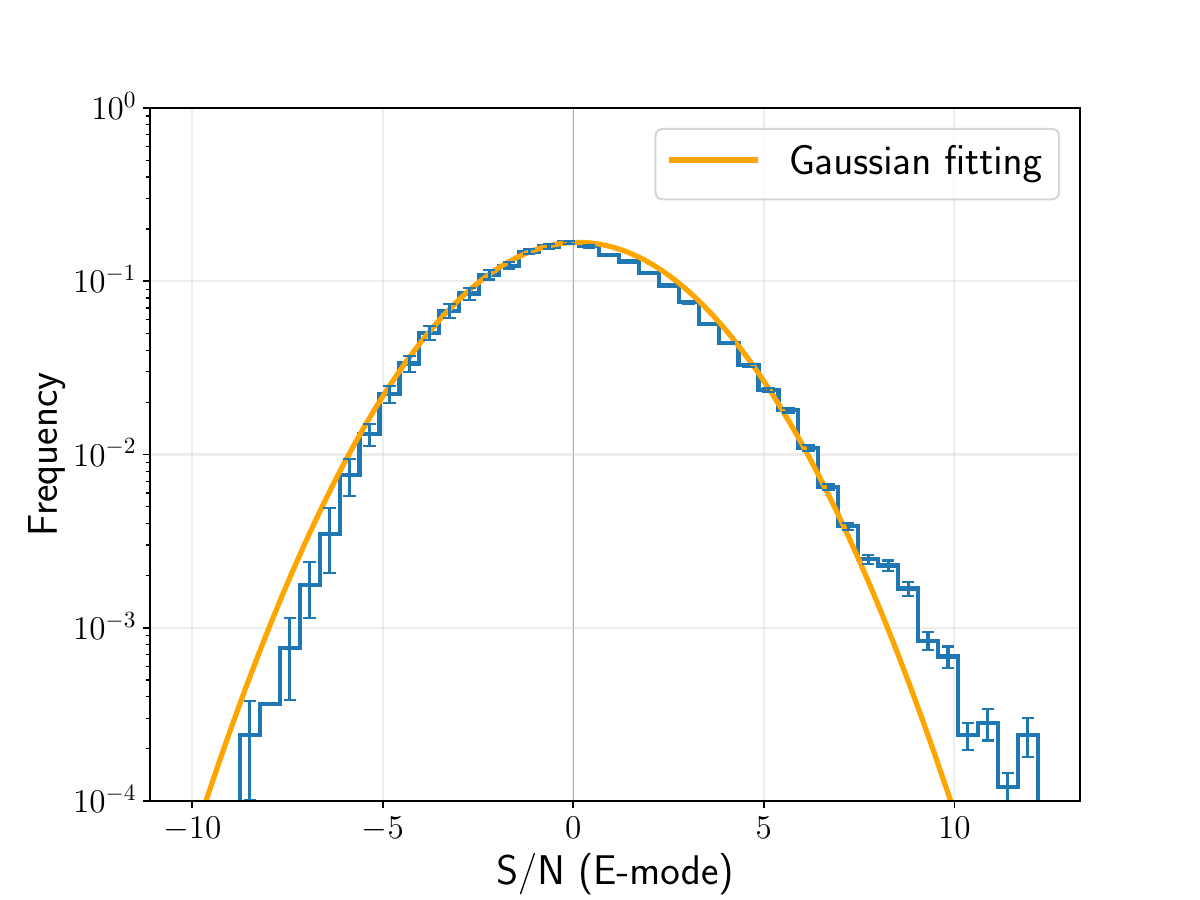}
    \includegraphics[width=\columnwidth]{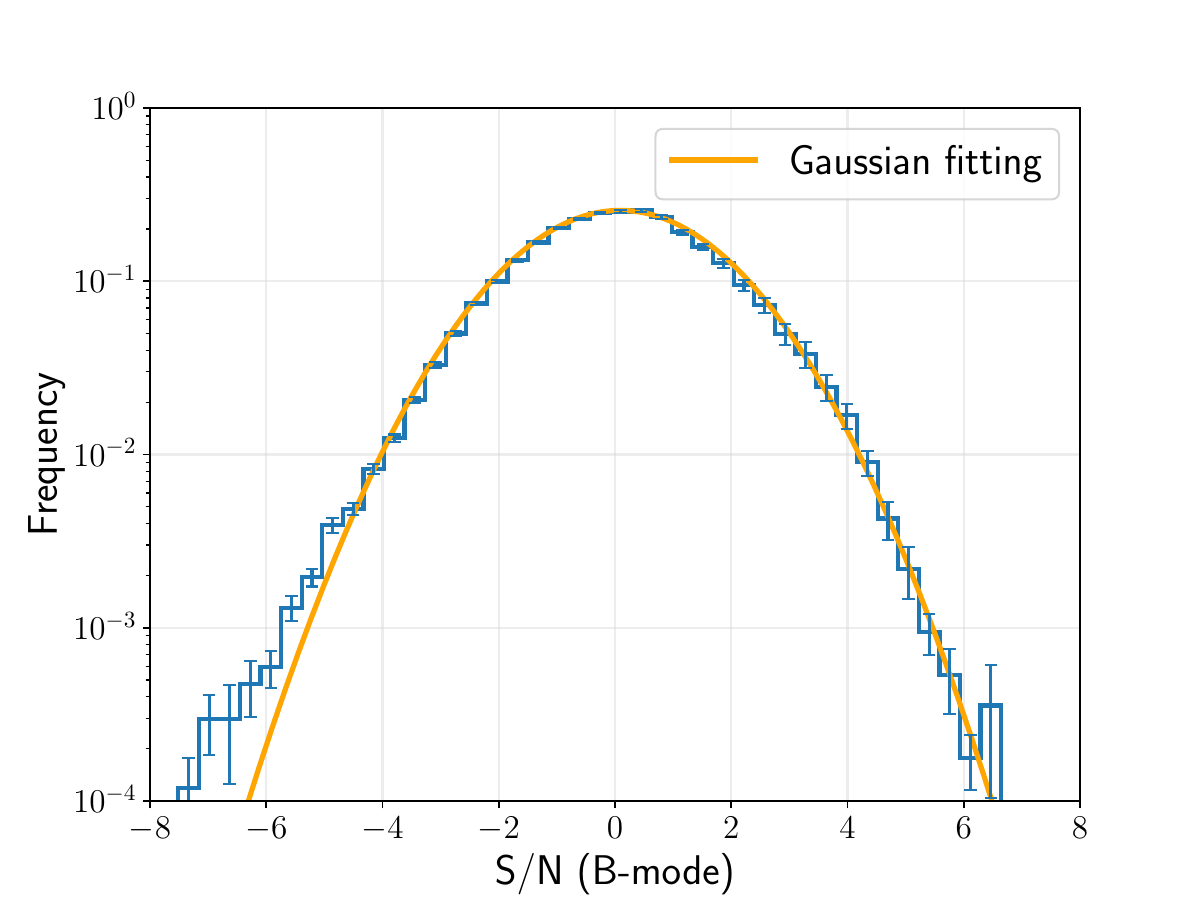}
    \caption{Probability distribution functions of E-mode (\textit{upper})
    and B-mode (\textit{lower}) S/N maps with errors from the noise map.
    The orange solid line shows the normal distribution with the sample mean 
    and standard deviation.}
    \label{fig:mass_map_hist}
\end{figure}

Figure~\ref{fig:mass_map_hist} shows the probability distribution functions (PDFs) of the 
S/N map of E-mode and B-mode convergence, where the E-mode convergence map corresponds to the mass map.
For the B-mode convergence map, the PDF is well approximated as the normal
distribution across all ranges. While the width of the best-fitting normal distribution is slightly larger than unity, this feature is expected when the map is dominated by the shape noise. Such broadening of the B-mode PDF is also seen in the analysis of mock weak lensing shape catalogues due to the boundary effect that partly mixes E- and B-mode signals \citep{Li2022}.
The slight deviation of the B-mode PDF from the Gaussian distribution at the lower tail ($\mathrm{S/N} \simeq -6$) is due to the residual systematics in weak lensing shape measurements and the similar feature is also found in \citet{Li2022}.
We find that the E-mode PDF near the peak is much broader than the B-mode PDF, indicating that the weak lensing signal is clearly detected. Furthermore, the PDF is highly skewed at $\mathrm{S/N} \simeq 10$, corresponding to weak lensing peaks created by massive structures
such as galaxy clusters.

\subsection{Three-dimensional mass mapping}
\label{sec:3D_mass_mapping}
The weak lensing mass map is the density distribution convolved with
the lensing kernel along the line-of-sight direction, and hence,
the line-of-sight information of the density structure
is not accessible due to the projection.
On the other hand, the three-dimensional structure of the density distribution
can be inferred with \textit{lensing tomography} \citep{Hu1999} technique,
where the source galaxy sample is divided into subsamples
according to the photometric, redshifts and mass maps are created for individual subsamples.
We can further push this idea to reconstruct the line-of-sight structures since the resultant mass maps have different lensing kernels \citep{Hu2002}.

For the reconstruction of three-dimensional mass maps, we follow the methodology
presented in \citet{Simon2009} and \citet{Oguri2018}.
Specifically, we consider convergence fields
for source galaxies with redshifts in the ranges of
$[z_{i, \mathrm{min}}, z_{i, \mathrm{max}}]$,
where the subscript $i = 1, \ldots, N_\kappa$ denotes the source redshift bin
and $N_\kappa$ is the number of the source redshift bins.
In this paper, we define the source redshift bin as the linear spacing with respect to the comoving distance:
\begin{align}
    \chi (z_{i, \mathrm{min}}) &= (300 + 300 \times i) \, \hMpc , \\
    \chi (z_{i, \mathrm{max}}) &= \chi (z_{i, \mathrm{min}}) + 300 \, \hMpc ,
\end{align}
with $N_\kappa = 14$.
The convergence $\kappa_i$ at the $i$-th source redshift bin
is expressed as the weighted sum of density contrasts in lens redshift bins. We consider density contrasts with redshifts in the range of
$[z_{j, \mathrm{min}}, z_{j, \mathrm{max}}]$,
where the subscript $j = 1, \ldots, N_\delta$ denotes the lens redshift bin. In this paper, we define the lens redshift bin as the linear spacing with respect to the comoving distance:
\begin{align}
    \chi (z_{j, \mathrm{min}}) &= (100 + 200 \times j) \, \hMpc , \\
    \chi (z_{j, \mathrm{max}}) &= \chi (z_{j, \mathrm{min}}) + 200 \, \hMpc ,
\end{align}
with $N_\delta = 10$. Using this expression of the three-dimensional density field, the convergence $\kappa_i$ at the $i$-th source redshift bin is written as
\begin{align}
    \kappa_i &\approx \sum_{z_j < z_i} \left[ \int_{z_{j, \mathrm{min}}}^{z_{j, \mathrm{max}}}
    \frac{\bar{\rho}(z)}{H(z) (1+z) \Sigma_{\mathrm{crit}, i} (z)} \mathrm{d} z \right] \delta_j
    \nonumber \\
    &\equiv \sum_{z_j < z_i} Q_{ij} \delta_j ,
    \label{eq:3D_delta}
\end{align}
where $H(z)$ is the Hubble parameter.
The summation runs over bins that satisfy $z_j < z_i$ because only the foreground structure
contributes to the convergence. To put it another way, we set $Q_{ij} = 0$ if $z_j > z_i$.
The critical surface mass density $\Sigma_{\mathrm{crit}, i}$ for $i$-th source redshift bin is approximated as
\begin{equation}
    \Sigma_{\mathrm{crit}, i}^{-1} (z) \approx
    \frac{4\pi G}{c^2} f_K(\chi(z))
    \frac{f_K(\chi(z)-\chi(\bar{z}_i))}{f_K(\chi(\bar{z}_i))},
    \label{eq:sigma_crit}
\end{equation}
where $\bar{z} = (z_{i, \mathrm{min}} + z_{i, \mathrm{max}}) / 2$
is the mean redshift.

In principle, the three-dimensional density distribution $\delta_j$ can be obtained
by inverting the linear equation Eq.~\eqref{eq:3D_delta}.
However, this operation is numerically unstable in general,
and thus, we employ the Wiener filtering to reduce the noise in Fourier space.
To this end, we compute the power spectrum signal $S_{lm}$
and noise $N_{lm}$ for $l$-th and $m$-th redshift bins:
\begin{align}
    S_{lm} &= \delta^\mathrm{K}_{lm} \frac{1}{(\Delta \chi_l)^2}
    \int_{z_{l,\mathrm{min}}}^{z_{l,\mathrm{max}}}
    \frac{c \mathrm{d}z}{H(z)} \frac{1}{ \chi^2 (z)}
    P_\mathrm{m} (k = \ell / \chi, z), \\
    N_{lm} &= \delta^\mathrm{K}_{lm} \frac{\sigma_e^2}{\bar{n}_l},
\end{align}
where $\delta^\mathrm{K}_{lm}$ is the Kronecker delta,
$\Delta \chi_l \approx c/H(\bar{z}_l) (z_{l,\mathrm{max}} - z_{l,\mathrm{min}})$ is the width of comoving distance
for the $l$-th redshift bin,
$P_\mathrm{m} (k, z)$ is the non-linear matter power spectrum
computed with \textit{halofit} prescription \citep{Smith2003}
with the updated parameters \citep{Takahashi2012},
$\sigma_e\approx 0.35$ is the root mean square of the ellipticity computed directly from the weak lensing shape catalogue,
and $\bar{n}_k$ is the mean number density of source galaxies
in $k$-th redshift bin, which is again computed directly from the weak lensing shape catalogue.
The reconstructed three-dimensional density field in harmonic space
is given by the minimum-variance estimator:
\begin{equation}
    \tilde{\bm{\delta}}_{\ell m}
    = \tilde{W} (\ell) D (\ell)
    \left[ \alpha S^{-1} + Q^\mathrm{T} N^{-1} Q \right]^{-1} 
    Q^\mathrm{T} N^{-1} \tilde{\bm{\gamma}}_{\ell m} ,
\end{equation}
where $\tilde{W}(\ell)$ is the Gaussian filter in harmonic space,
$D(\ell) \equiv - \sqrt{\ell (\ell + 1)/((\ell - 1) (\ell + 2))}$ (Eq.~\ref{eq:kappa_sht}),
and $\tilde{\bm{\gamma}}_{\ell m} \equiv (\tilde{\gamma}_{1, \ell m}, \ldots, \tilde{\gamma}_{N_z, \ell m})$
is the shear field in harmonic space binned with the redshifts.
We have introduced the parameter $\alpha$
which controls the regularization with the signal power spectrum and we adopt $\alpha = 0.03$ following \citet{Oguri2018}.
Finally, the three-dimensional density field $\delta_i$
can be obtained by applying the inverse spherical transformation to $\tilde{\bm{\delta}}_{\ell m} \equiv (\tilde{\delta}_{1, \ell m}, \ldots, \tilde{\delta}_{N_\kappa, \ell m})$.

\subsection{Subaru Hyper Suprime-Cam weak lensing shape catalogue}
\label{sec:HSC_shape_catalogue}
The Hyper Suprime-Cam (HSC) installed on the Subaru Telescope is a wide-field imaging camera
\citep{Miyazaki2018}, with the field-of-view of $1.77 \, \mathrm{deg}^2$.
In this work, we use the Year 3 (Y3) weak lensing shape catalogue based on
the $i$-band coadded images from the wide layer of the HSC Strategic Survey Program
\citep[HSC-SSP;][]{Aihara2018a,Aihara2018b,Aihara2019,Aihara2022}
taken from March 2014 to April 2019.
The full details of the production pipeline
of the Y3 weak lensing shape catalogue are described in \citet{Li2022}.
Here, we overview the basics of the shape catalogue.
The galaxy ellipticities $\bm{e}$ are estimated with the \texttt{GalSim} code \citep{Rowe2014}
based on the re-Gaussianization method \citep{Hirata2003}:
\begin{equation}
\bm{e} = \frac{1-(b/a)^2}{1+(b/a)^2} (\cos 2\phi, \sin 2\phi),
\end{equation}
where $b/a$ is the major-to-minor axis ratio of the galaxy isophotes and
$\phi$ is the polar angle measured from the major axis.
The re-Guassianization method is subject to the shear estimation bias
and these biases are corrected by introducing the multiplicative bias $m$
and additive biases $c_1$ and $c_2$.
These bias terms are calibrated with image simulations as a function of galaxy properties,
e.g., signal-to-noise ratio and redshift.

\begin{figure*}
\includegraphics[width=\textwidth]{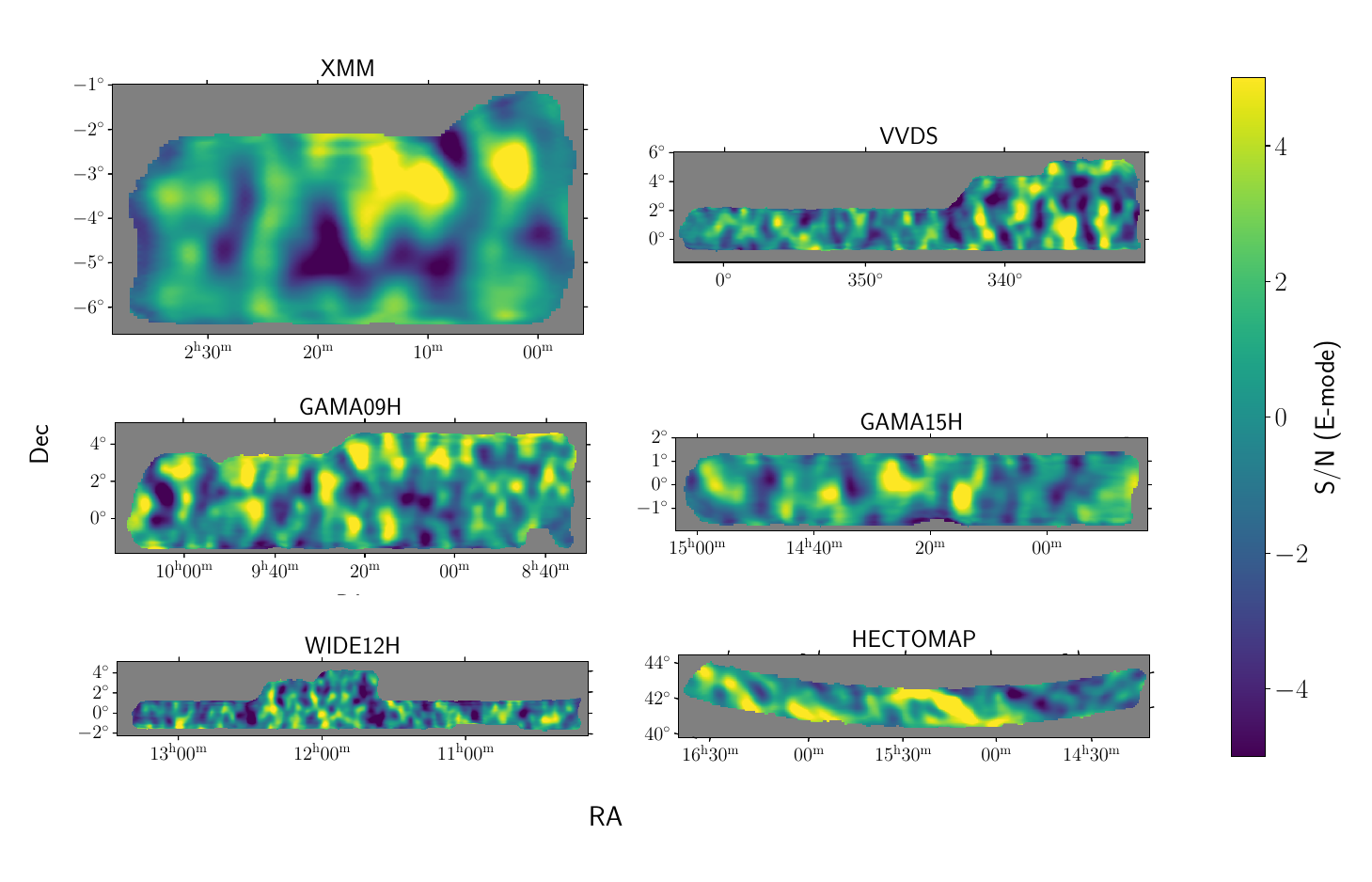}
\caption{The S/N maps of E-mode convergence fields for six HSC survey patches:
XMM, VVDS, GAMA09H, GAMA15H, WIDE12H, HECTOMAP.}
\label{fig:HSC_survey}
\end{figure*}

We only consider the area where the data in all five HSC broadbands ($g,r,i,z,y$) reach the full-depth and the error of the point spread function is small.
The total area of the catalogues is $433.48\, \mathrm{deg}^2$,
which are divided into six patches: XMM, VVDS, GAMA09H, GAMA15H, WIDE12H, and HECTOMAP.
Figure~\ref{fig:HSC_survey} shows the S/N map derived from the HSC Y3 shape catalogue.

The photometric redshifts of source galaxies in Y3 shape catalogues
are estimated with various approaches \citep[for HSC Y1 weak lensing analysis, see][]{Tanaka2018}.
We adopt \texttt{dNNz} method (Nishizawa, et al., in prep.),
which utilizes neural networks
with inputs of \texttt{cmodel} magnitudes, size, and point spread function
matched aperture magnitudes in all five HSC bands.
For each galaxy, the probability distribution function of the redshift
in the range of $0 < z < 7$ with $100$ bins is computed.
The performance is evaluated with the test sample,
which consists of $10$ per cent of the whole sample
and is not used in the training process;
the bias is $< 10^{-4}$, the scatter is $3$ per cent,
and the outlier fraction is $< 10$ per cent.

\section{Photometric and spectroscopic galaxy catalogues}
\label{sec:catalogues}
To study the line-of-sight density structures
at the positions of troughs in detail, we utilize the three-dimensional galaxy distributions,
which are the biased tracers of the large-scale structures, adopting two different galaxy catalogues.
First, we employ the photometric LRG catalogue constructed
from the HSC photometric data.
Since only the photometric redshifts are available for the LRG catalogue,
the uncertainty of the line-of-sight distance of galaxies is relatively large. On the other hand, an advantage of the HSC photometric galaxy catalogue is that it uniformly covers the entire weak lensing mass map regions with the high galaxy number density out to the high redshift of $z \simeq 1.2$, 
enabling us to address the line-of-sight density structures
with less statistical noise.
Next, the spectroscopic galaxy samples from SDSS are used. 
This spectroscopic catalogue is complementary to the photometric LRG catalogue because the redshift is robustly determined with spectroscopy. Hence, the line-of-sight structure is not smeared by the redshift uncertainty. On the other hand, the spectroscopic catalogue contains fewer galaxies with a narrower range of redshifts ($0.1 < z < 0.8$).
Leveraging these two catalogues, we investigate the three-dimensional structures of high-significance troughs in weak lensing mass maps.

\subsection{Photometric galaxy catalogue}
\begin{figure}
  \includegraphics[width=\columnwidth]{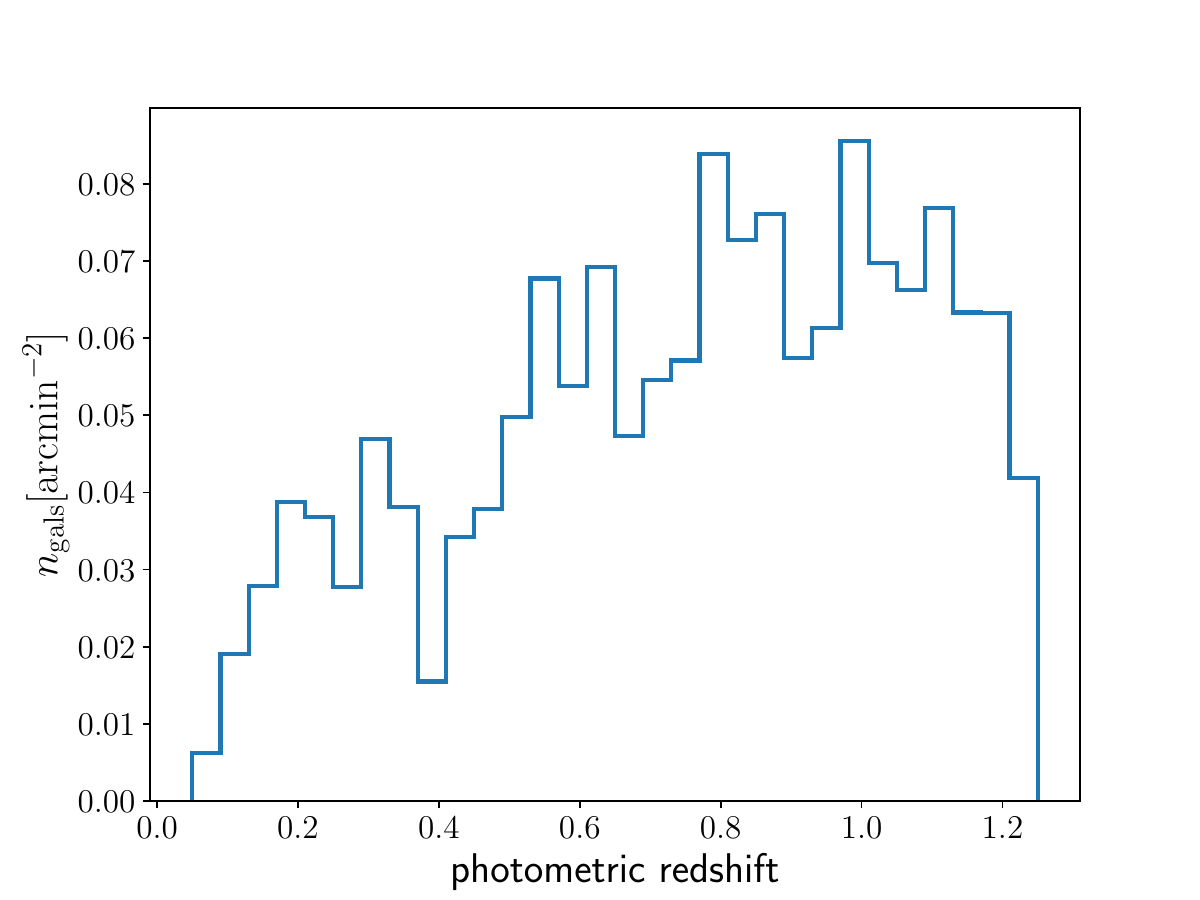}
  \caption{The redshift distribution of the number density of the HSC photometric LRGs.}
  \label{fig:pz_HSC}
\end{figure}

First, we employ the photometric LRG catalogue in HSC survey regions identified using the stellar population synthesis model fitting of galaxy colours used for Cluster-finding Algorithm based on Multi-band Identification of Red-sequence gAlaxies \citep[CAMIRA;][]{Oguri2014,Oguri2018a,Oguri2018}.
With the help of the deep multi-band imaging by HSC, the photometric LRG catalogue is constructed out to the redshift of $z\sim 1$ with redshift accuracy of $\Delta z / (1+z) \sim 0.03$ \citep{Oguri2018,Ishikawa2021}.
We use the photometric LRG catalogue
based on the latest internal HSC-SSP data that contains 5,479,879 LRGs in the redshift range of $0.05 < z < 1.25$.
The redshift distribution of the photometric LRGs is shown in Figure~\ref{fig:pz_HSC}.
Although the redshift accuracy of the photometric catalogue is not so high
as compared with the spectroscopic catalogue, its high number density is helpful for the robust investigation of the density structure in the line-of-sight direction
at the positions of weak lensing troughs.

\subsection{Spectroscopic galaxy catalogue}
\begin{figure}
  \includegraphics[width=\columnwidth]{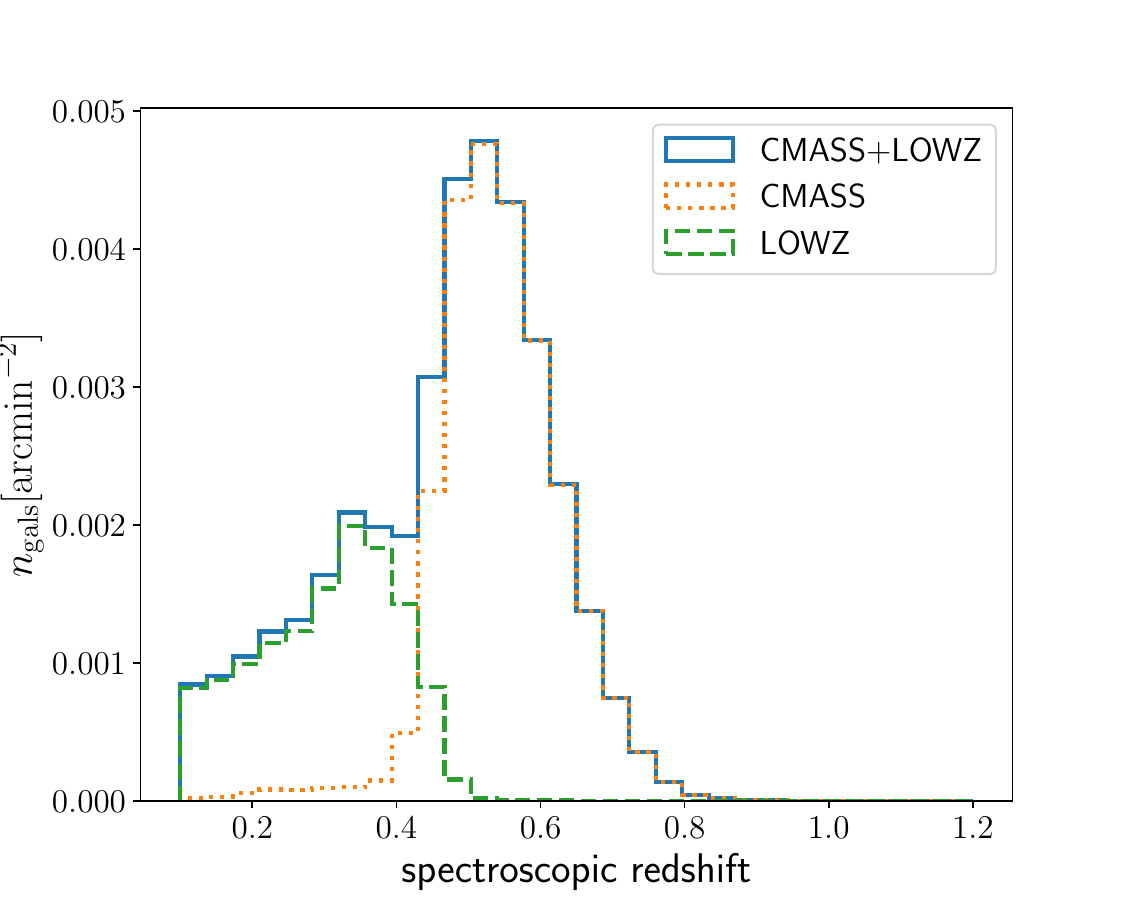}
  \caption{The redshift distributions of the number densities of
  LOWZ (\textit{green dashed}), CMASS (\textit{orange dotted}),
  and the sum of LOWZ and CMASS samples (\textit{blue solid})
  in SDSS DR12.}
  \label{fig:pz_SDSS}
\end{figure}

Next, we employ the spectroscopic galaxy catalogue
that provides highly accurate galaxy redshifts.
To this end, we use the galaxy catalogue of
SDSS-III Baryon Oscillation Spectroscopic Survey (BOSS) DR12 data \citep{Reid2016}.
In particular, CMASS and LOWZ galaxy samples are
employed to address the line-of-sight density structures.
These two catalogues are constructed based on the different
colour and magnitude cuts \citep{Eisenstein2001,Cannon2006}.
These selection criteria are designed so that
the redshift ranges of CMASS and LOWZ are $0.4 < z < 0.8$
and $z \lesssim 0.4$, respectively.
The effective areas of CMASS and LOWZ are $9376 \, \mathrm{deg}^2$
and $8337 \, \mathrm{deg}^2$, respectively.
We note that the HSC and BOSS survey regions are almost overlapped.
The total number of galaxies is 777,202 for CMASS
and 361,762 for LOWZ.
Figure~\ref{fig:pz_SDSS} shows the redshift distributions
of CMASS and LOWZ galaxy samples.
Although the number densities of CMASS and LOWZ samples
are substantially lower than the HSC photometric LRG sample,
the precise redshift determination of CMASS and LOWZ samples
complements the analysis with the HSC photometric LRG sample.

\section{Results}
\label{sec:results}
In this Section, we present troughs identified in mass maps
constructed from the HSC Y3 shape catalogue.
We then measure the radial profiles of the convergence for representative troughs
to see how the underdense regions extend in the sky.
Next, we measure the redshift distributions of photometric LRGs and CMASS/LOWZ galaxies
at the positions of selected troughs
to investigate whether the trough originates from single or multiple underdense regions along the line-of-sight.
Finally, we present the three-dimensional mass maps around the troughs 
and discuss their consistency with the galaxy distributions.

\subsection{Properties of identified troughs}
First, we present the S/N and positions of troughs identified in our analysis
in descending order of significance in Table~\ref{tab:trough_list}.
The possible origin of each trough, i.e., a single void or alignment of multiple voids along the line-of-sight, can be discriminated by the redshift distribution of the photometric LRGs and the CMASS/LOWZ galaxies
at the position of the trough,
which will be shown in Section~\ref{sec:los_structure_galaxy}.
Some troughs are located close to the edge of the survey region.
The troughs in the edge regions may be spurious
because the mass map is noisier due to the mixing of E- and B-mode near the edge.
We classify troughs which are located less than $60 \, \mathrm{arcmin}$ from the nearest masked region  as ``Edge origin''.

The redshift distribution describes the number density
at the trough position divided by the mean number density
as a function of redshift, to which
we refer to the line-of-sight number density contrast.
Specifically, it is defined as $n_{\mathrm{trough}}/n_\mathrm{mean}$,
where $n_{\mathrm{trough}}(z_i)$ is the number of galaxies
within the $z_i$ bin and the top-hat aperture radius
of $10 \,\mathrm{arcmin}$ ($50 \,\mathrm{arcmin}$)
for HSC photometric LRGs
(CMASS/LOWZ galaxies\footnote{The aperture radius of $50 \, \mathrm{arcmin}$
for CMASS/LOWZ galaxies is much larger than the fibre collision scale $\sim 1\, \mathrm{arcmin}$.})
from the centre of the trough divided by the aperture area.
The aperture radii for HSC and SDSS are determined so that the similar number of galaxies are included within the aperture,
which yields the similar Poisson noise.

The mean number density $n_\mathrm{mean}(z_i)$ for HSC photometric LRGs is
calculated as the number of galaxies
in each redshift bin divided by the total survey area.
For CMASS/LOWZ galaxies, the random catalogues are
provided in SDSS DR12.
We employ the random catalogues to compute
the mean number density to incorporate the survey mask effect
and completeness of the CMASS/LOWZ galaxies.
The redshift bins are equally spaced in the range of $0.1 < z < 1.2$
with $10$ bins.
The error is estimated as Poisson noise.

We classify troughs into two different classes according to their origin:
``single-void \textit{candidate}'' or ``multiple-voids''.
We define the single-void origin as the trough region
where both LRG and CMASS/LOWZ catalogues exhibit
the single underdense regions deeper than $4\sigma$ with respect to the mean
in the line-of-sight redshift distribution
and multiple-void origin as those having multiple underdense regions, respectively.
Here, we regard the single-void origin troughs as \textit{candidates}
because the line-of-sight density structure probed by galaxy samples employed in this analysis
has large uncertainties.
To identify the void origins,
we mainly use the photometric HSC catalogue because
the CMASS/LOWZ catalogue does not fully cover the HSC Y3 weak lensing survey footprint.
We use CMASS/LOWZ galaxies for cross-checking in the region where they are available.
Note that the LOWZ and CMASS survey regions are not fully overlapped, and as a result, the number densities of spectroscopic galaxies can be extremely small for some of the troughs.

Excluding troughs near the edge, there are four troughs with $|\mathrm{S/N}| > 7$, all of which turn out to be the multiple-voids origin. By lowering the $|\mathrm{S/N}|$ threshold, we find 15 troughs with $|\mathrm{S/N}| > 5.7$, two of which are single-void origin candidates.

\begin{table*}
\caption{Troughs with $|\mathrm{S}/\mathrm{N}| > 5.7$ in the mass maps constructed
from the HSC Y3 weak lensing shape catalogue.
The label is numbered in descending order with respect to $|\mathrm{S}/\mathrm{N}|$.
Note that $\mathrm{S}/\mathrm{N}$ is negative in general because
the convergence at a trough is negative.
The possible source of each trough is judged by visual inspections
of the redshift distribution of photometric LRGs and CMASS/LOWZ galaxies
at the position of each trough.}
    \begin{tabular}{cccccc}\hline
        Label & $|\mathrm{S}/\mathrm{N}|$ & RA & Dec & Source & Redshifts of source voids \\
        \hline\hline
        T1	&	$8.77$	&	$12^\mathrm{h}29^\mathrm{m}53^\mathrm{s}$   &	$-00\degr35\arcmin49\arcsec$ & Multiple voids & 0.6, 1.0 \\
        T2	&	$7.83$	&	$10^\mathrm{h}04^\mathrm{m}41^\mathrm{s}$	&	$+01\degr25\arcmin03\arcsec$ & Multiple	voids & 0.2, 0.5 \\
        T3	&	$7.21$	&	$10^\mathrm{h}57^\mathrm{m}04^\mathrm{s}$	&	$-00\degr08\arcmin57\arcsec$ & Multiple	voids & 0.6, 0.7 \\
        T4	&	$7.15$	&	$11^\mathrm{h}36^\mathrm{m}06^\mathrm{s}$	&	$-00\degr31\arcmin20\arcsec$ & Multiple	voids & 0.7, 1.0 \\
        T5	&	$7.01$	&	$22^\mathrm{h}53^\mathrm{m}33^\mathrm{s}$	&	$+02\degr44\arcmin12\arcsec$ & Edge & --- \\
        T6	&	$6.92$	&	$14^\mathrm{h}54^\mathrm{m}07^\mathrm{s}$	&	$+44\degr06\arcmin08\arcsec$ & Multiple	voids & 0.1, 0.3 \\
        T7	&	$6.79$	&	$22^\mathrm{h}43^\mathrm{m}22^\mathrm{s}$   &	$+02\degr54\arcmin 39\arcsec$ & Single void (candidate) & 0.2 \\
        T8	&	$6.67$	&	$02^\mathrm{h}08^\mathrm{m}19^\mathrm{s}$	&	$-02\degr14\arcmin 19\arcsec$ & Edge & --- \\
        T9	&	$6.67$	&	$02^\mathrm{h}18^\mathrm{m}52^\mathrm{s}$	&	$-04\degr55\arcmin48\arcsec$ & Multiple voids & 0.3, 0.6 \\
        T10	&	$6.42$	&	$12^\mathrm{h}00^\mathrm{m}00^\mathrm{s}$	&	$+04\degr15\arcmin 23\arcsec$ & Edge & --- \\
        T11	&	$6.37$	&	$11^\mathrm{h}42^\mathrm{m}46^\mathrm{s}$	&	$+02\degr14\arcmin 19\arcsec$ & Multiple voids & 0.2, 0.5 \\
        T12	&	$6.36$	&	$10^\mathrm{h}38^\mathrm{m}05^\mathrm{s}$	&	$+00\degr00\arcmin 00\arcsec$ & Multiple voids & 0.6, 1.0 \\
        T13	&	$6.24$	&	$09^\mathrm{h}06^\mathrm{m}41^\mathrm{s}$	&	$-01\degr02\arcmin 40\arcsec$ & Edge & --- \\
        T14	&	$6.24$	&	$14^\mathrm{h}16^\mathrm{m}45^\mathrm{s}$	&	$-01\degr29\arcmin 32\arcsec$ & Edge & --- \\
        T15	&	$6.05$	&	$22^\mathrm{h}54^\mathrm{m}37^\mathrm{s}$	&	$+00\degr22\arcmin 23\arcsec$ & Multiple voids  & 0.4, 0.6 \\
        T16	&	$6.05$	&	$12^\mathrm{h}00^\mathrm{m}21^\mathrm{s}$	&	$+04\degr28\arcmin 51\arcsec$ & Edge & --- \\
        T17	&	$6.01$	&	$22^\mathrm{h}03^\mathrm{m}38^\mathrm{s}$	&	$+01\degr38\arcmin 29\arcsec$ & Edge & --- \\
        T18	&	$5.96$	&	$09^\mathrm{h}10^\mathrm{m}54^\mathrm{s}$	&	$+01\degr02\arcmin 40\arcsec$ & Multiple voids & 0.1, 0.6, 0.7, 0.8 \\
        T19	&	$5.92$	&	$12^\mathrm{h}17^\mathrm{m}14^\mathrm{s}$	&	$+02\degr32\arcmin 15\arcsec$ & Multiple voids & 0.2, 0.3 \\
        T20	&	$5.89$	&	$11^\mathrm{h}16^\mathrm{m}45^\mathrm{s}$	&	$+00\degr53\arcmin 43\arcsec$ & Multiple voids & 0.2, 0.6 \\
        T21	&	$5.84$	&	$02^\mathrm{h}18^\mathrm{m}52^\mathrm{s}$	&	$-04\degr19\arcmin 52\arcsec$ & Multiple voids & 0.3, 0.6 \\
        T22	&	$5.73$	&	$11^\mathrm{h}03^\mathrm{m}03^\mathrm{s}$	&	$+01\degr16\arcmin 06\arcsec$ & Edge & --- \\
        T23	&	$5.73$	&	$22^\mathrm{h}19^\mathrm{m}27^\mathrm{s}$	&	$+04\degr06\arcmin 24\arcsec$ & Single void  (candidate) & 0.3 \\
        \hline
    \end{tabular}
    \label{tab:trough_list}
\end{table*}

\begin{figure*}
    \includegraphics[width=2\columnwidth]{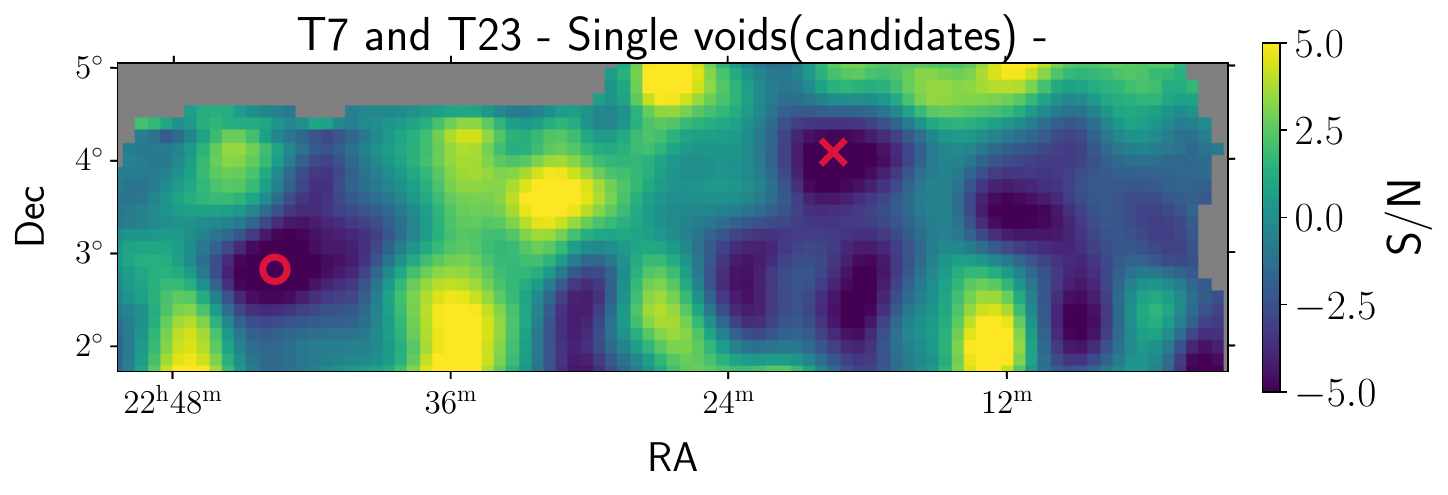}
    \caption{The S/N maps around the troughs T7 (\textit{circle symbol})
    and T23 (\textit{cross symbol}), which are single-void origin candidates.}
    \label{fig:mass_map_single_void}
\end{figure*}

\begin{figure}
    \includegraphics[width=\columnwidth]{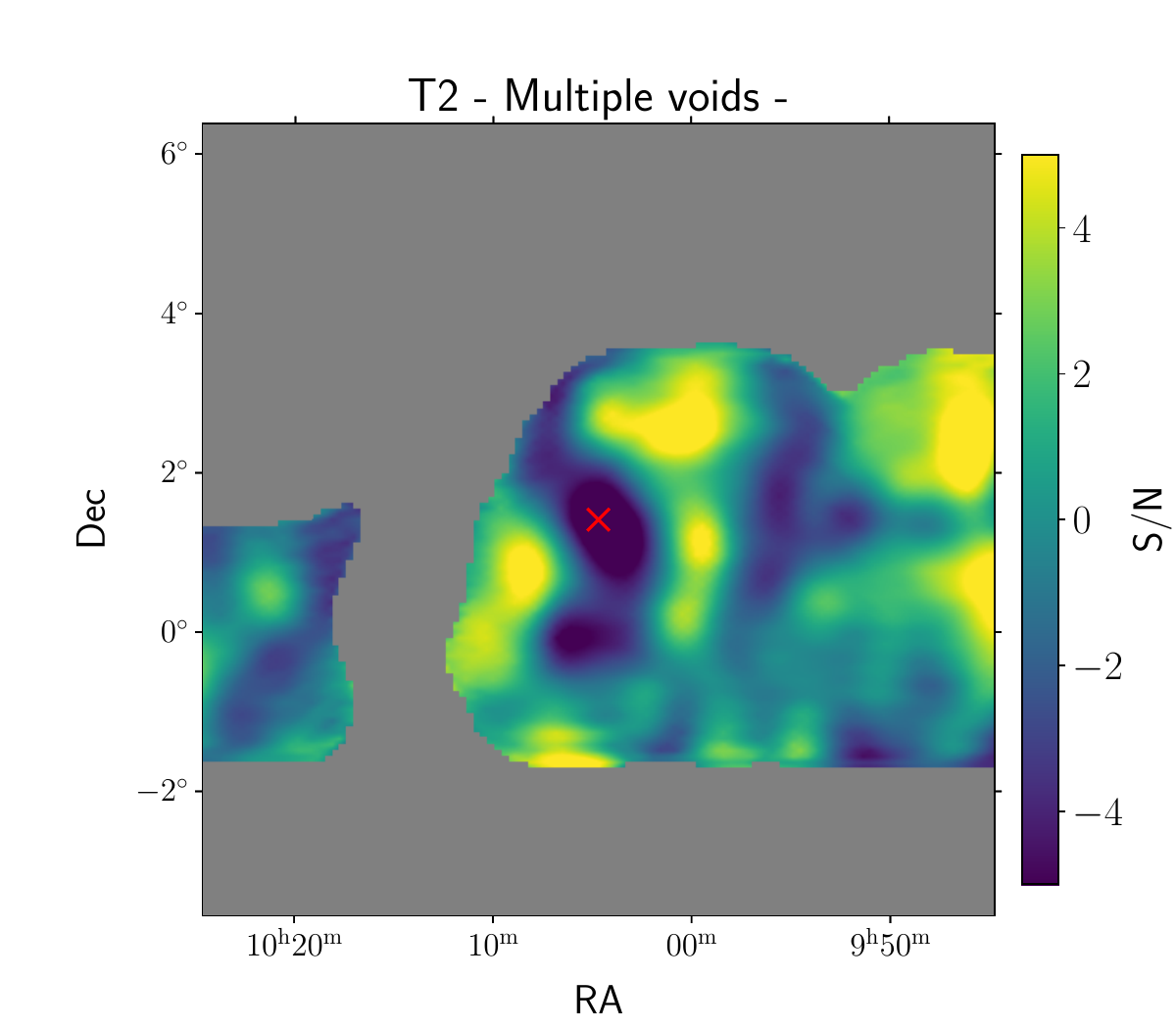}
    \includegraphics[width=\columnwidth]{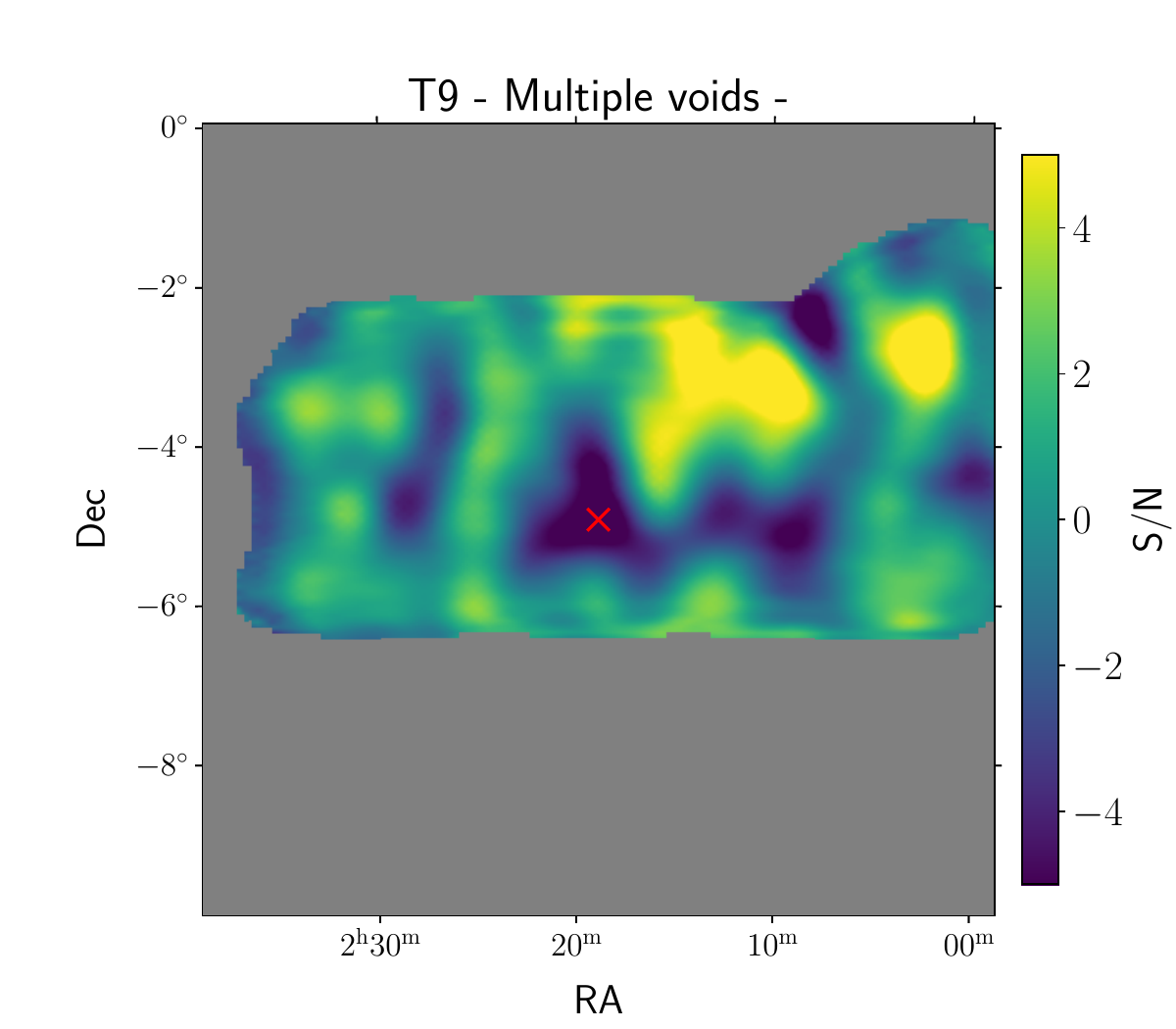}
    \caption{Similar to Figure. \ref{fig:mass_map_single_void}, but for T2 (\textit{upper}) and T9 (\textit{lower}).}
    \label{fig:mass_map_multiple_voids}
\end{figure}

In what follows, we present a line-of-sight property of the most prominent two single-void origin candidates, T7 and T23. For comparison, we also show the line-of-sight property of T2 and T9 as representative examples of multiple-voids origin troughs.
Figures~\ref{fig:mass_map_single_void} and \ref{fig:mass_map_multiple_voids} show the S/N maps around these troughs.
It is seen that these troughs are located far ($> 1\,\mathrm{deg}$) from the survey edge, and thus, they are not likely to be artefacts.

\subsection{Radial profile of the convergence around troughs}
\begin{figure}
    \includegraphics[width=\columnwidth]{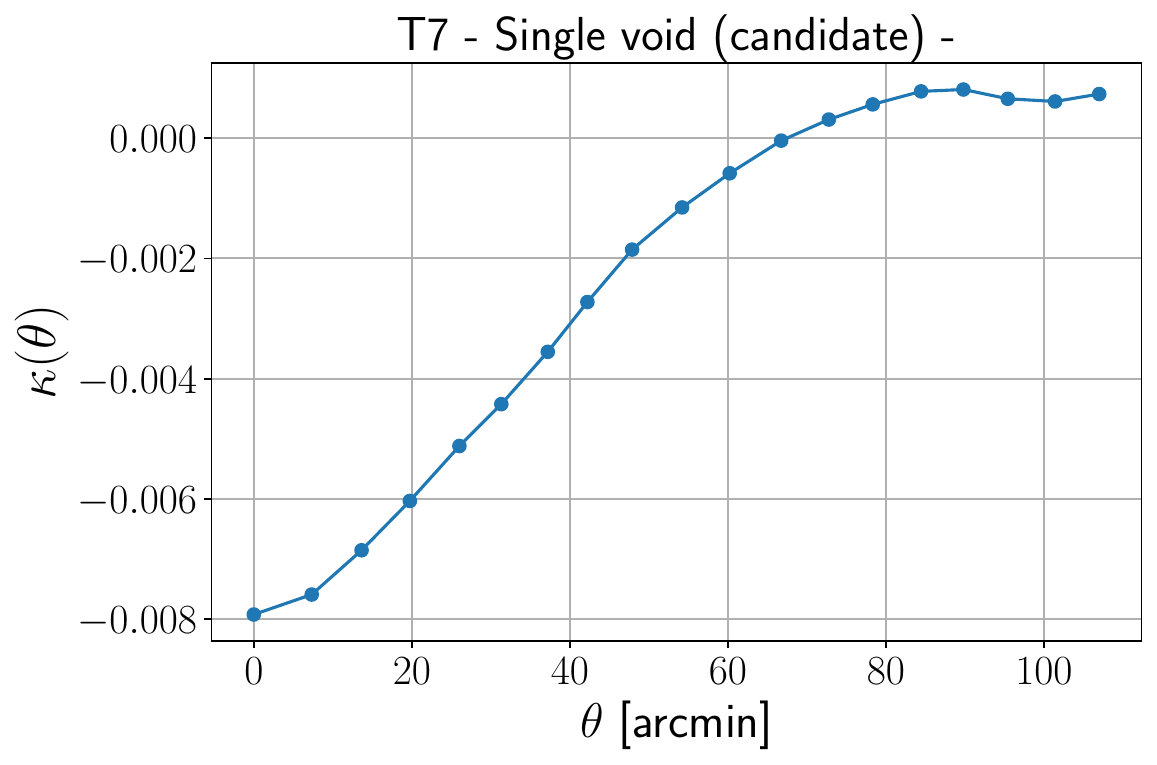}
    \includegraphics[width=\columnwidth]{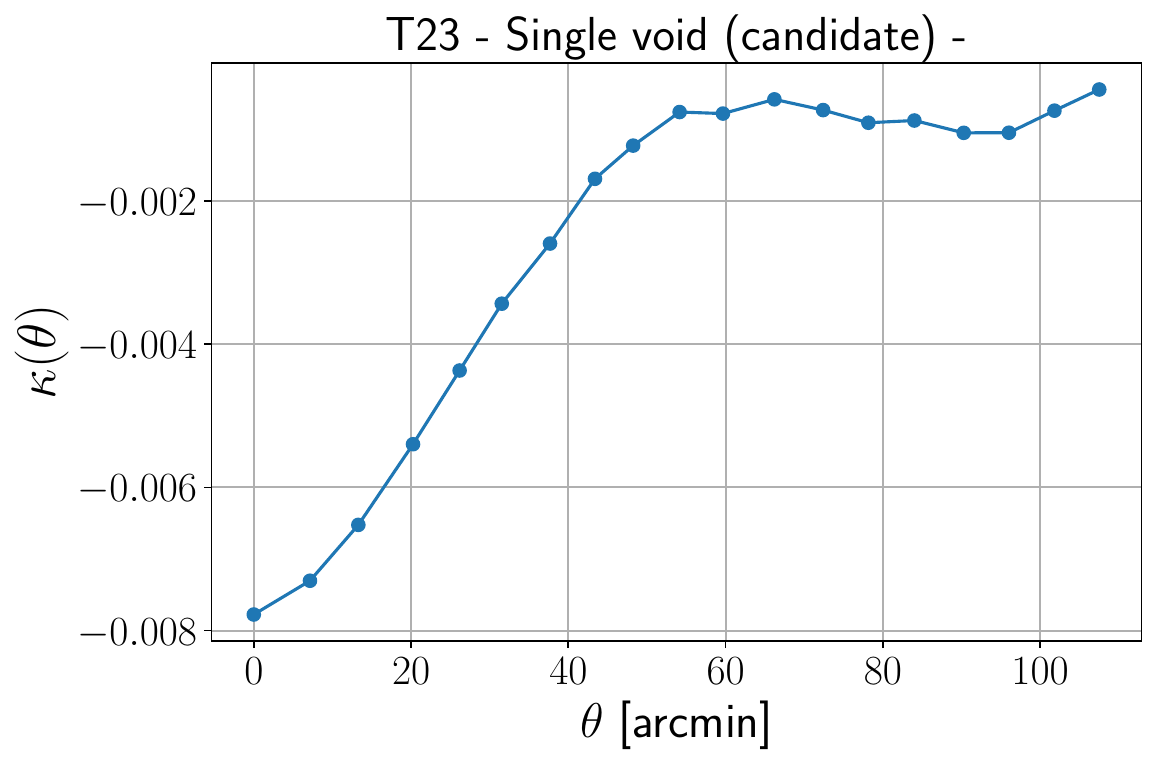}
    \caption{Radial profiles of the convergence centred
    at T7 (\textit{upper}) and T23 (\textit{lower}).}
    \label{fig:profile_single_void}
\end{figure}

\begin{figure}
    \includegraphics[width=\columnwidth]{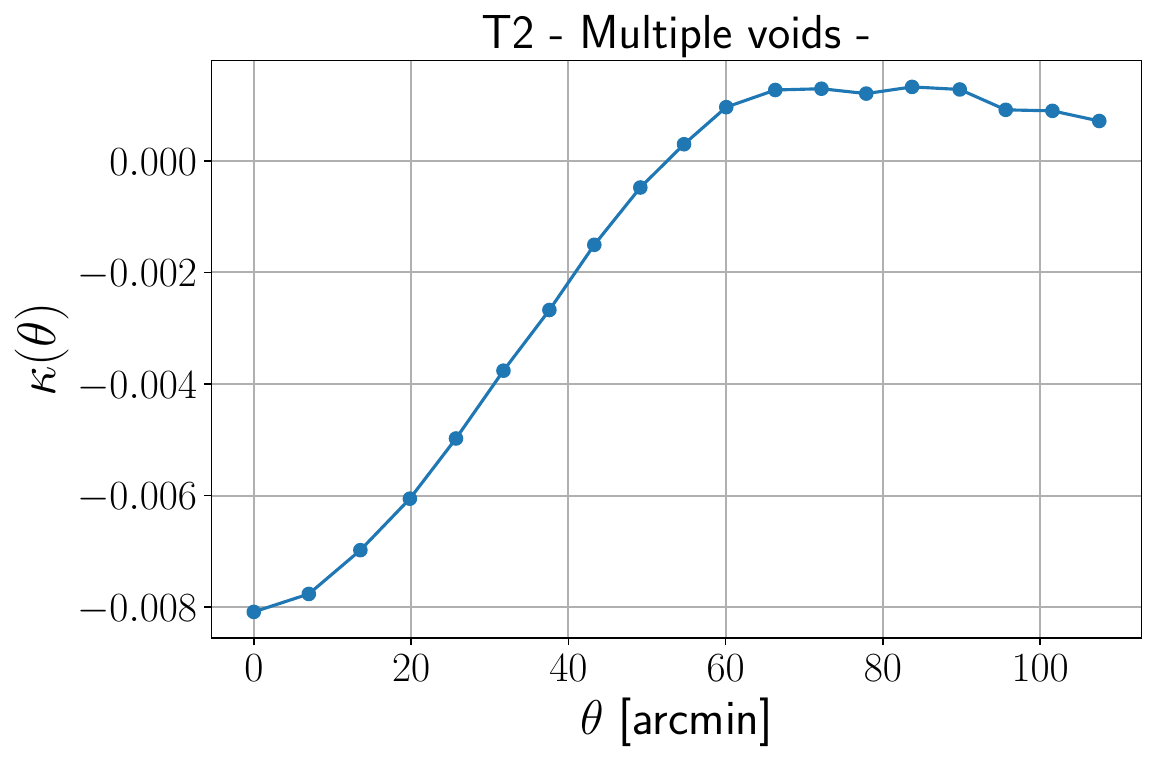}
    \includegraphics[width=\columnwidth]{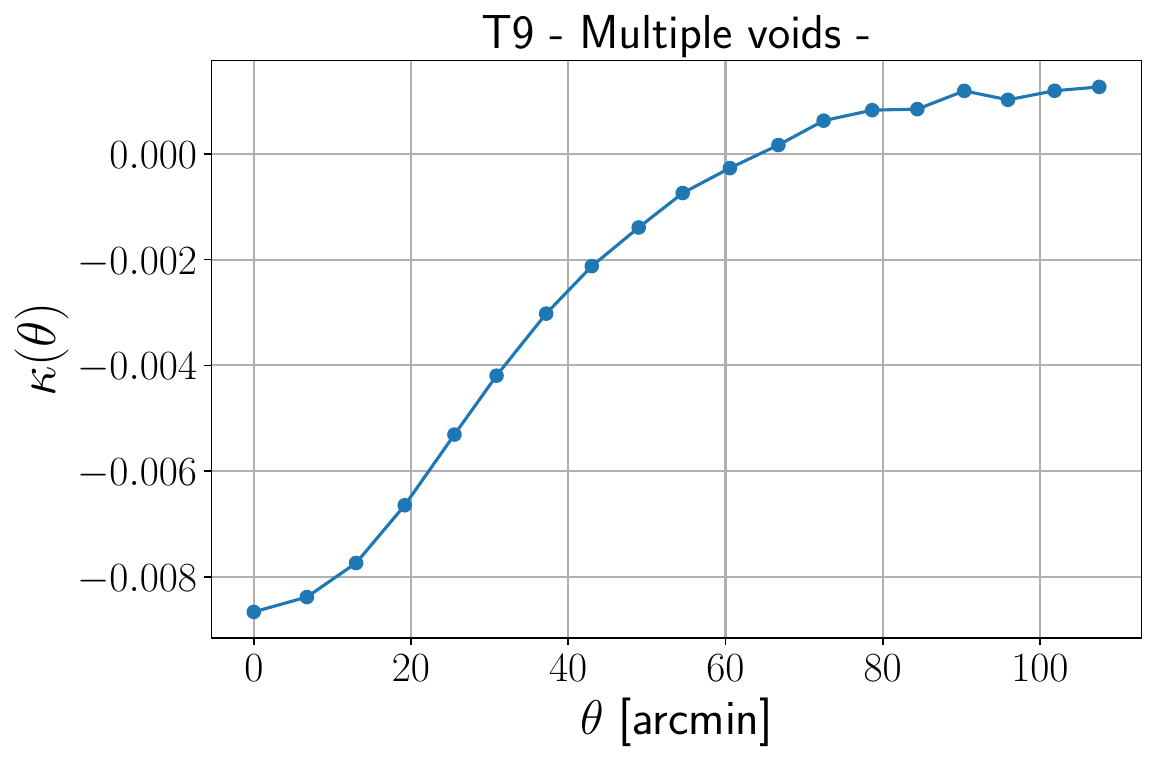}
    \caption{Similar to Figure~\ref{fig:profile_single_void}, but for T2 (\textit{upper}) and T9 (\textit{lower}).}
    \label{fig:profile_multiple_voids}
\end{figure}

We measure radial profiles of the convergence centred at the troughs.
The radial profile is derived by computing the mean of convergence in the annuli,
which are equally spaced in the range of $0 < \theta [\mathrm{arcmin} ]< 110$ with $19$ bins.
Figures~\ref{fig:profile_single_void} and \ref{fig:profile_multiple_voids} show the measured radial profiles.
The decrement of the convergence extends to $\simeq 1 \, \mathrm{deg}$,
which is larger than the smoothing scale of $40 \, \mathrm{arcmin}$ in FWHM,
implying that the troughs originate from spatially extended underdense regions.

As a sanity check, we measure the radial profiles from weak lensing mass maps with smoothing scales of $20$ and $60 \, \mathrm{arcmin}$. Their convergence profiles are similar for different smoothing scales, although the $|\mathrm{S/N}|$ at the centres are different for different smoothing scales because the noise maps depend on the smoothing scale.  The radii of voids for T7 and T23 are not sensitive to the smoothing scale, and the locations of the troughs are also similar with small shifts by one pixel, corresponding to $6.9 \, \mathrm{arcmin}$. Thus, we conclude that the identified troughs are of physical origin.

\subsection{Line-of-sight structure probed by galaxy distributions}
\label{sec:los_structure_galaxy}
To study the line-of-sight density structure of each trough, we utilize
the galaxy distribution, which is a biased tracer of the underlying dark matter distribution.
Specifically, we employ
HSC photometric LRGs and CMASS/LOWZ catalogues, which are described
in Section~\ref{sec:catalogues}, and visually inspect line-of-sight distributions of HSC photometric
LRGs and CMASS/LOWZ galaxies around the troughs.

\begin{figure*}
    \includegraphics[width=0.45\textwidth]{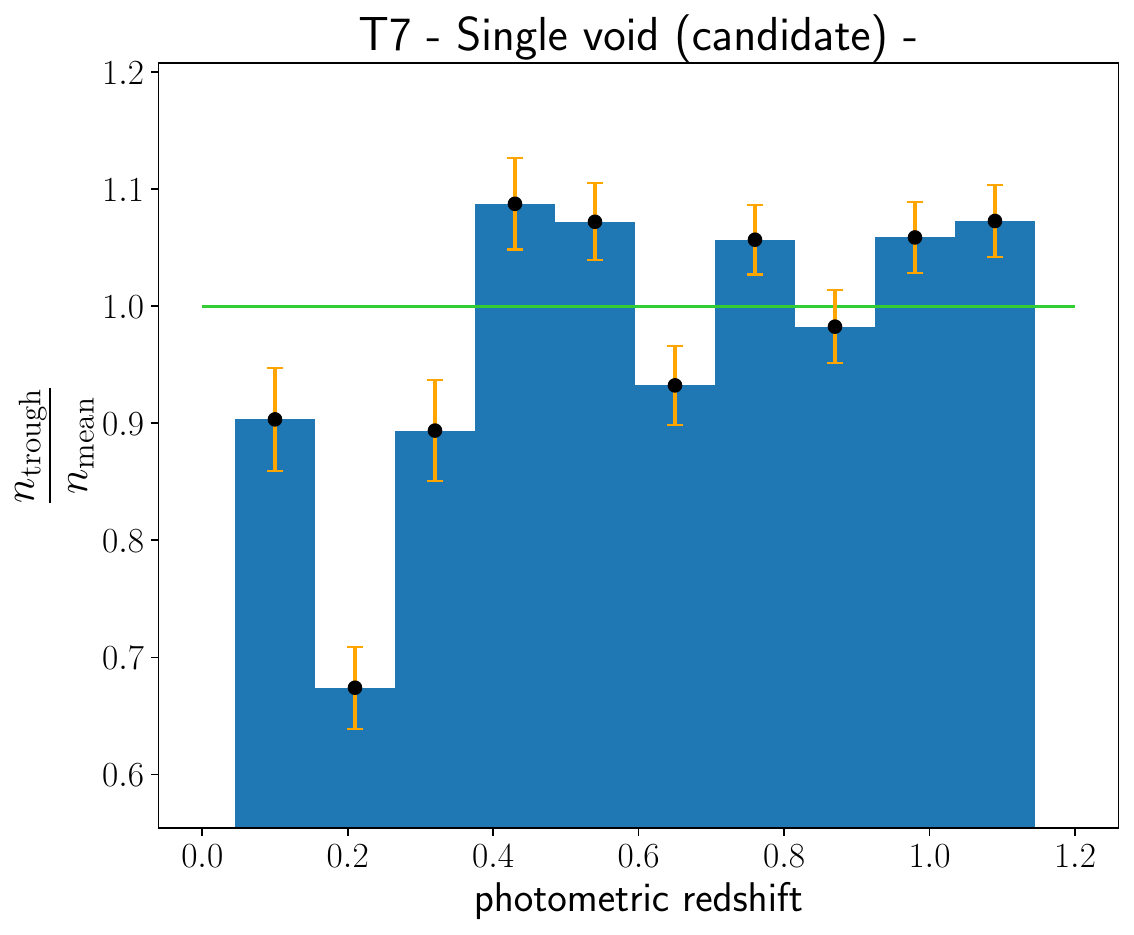}
    \includegraphics[width=0.45\textwidth]{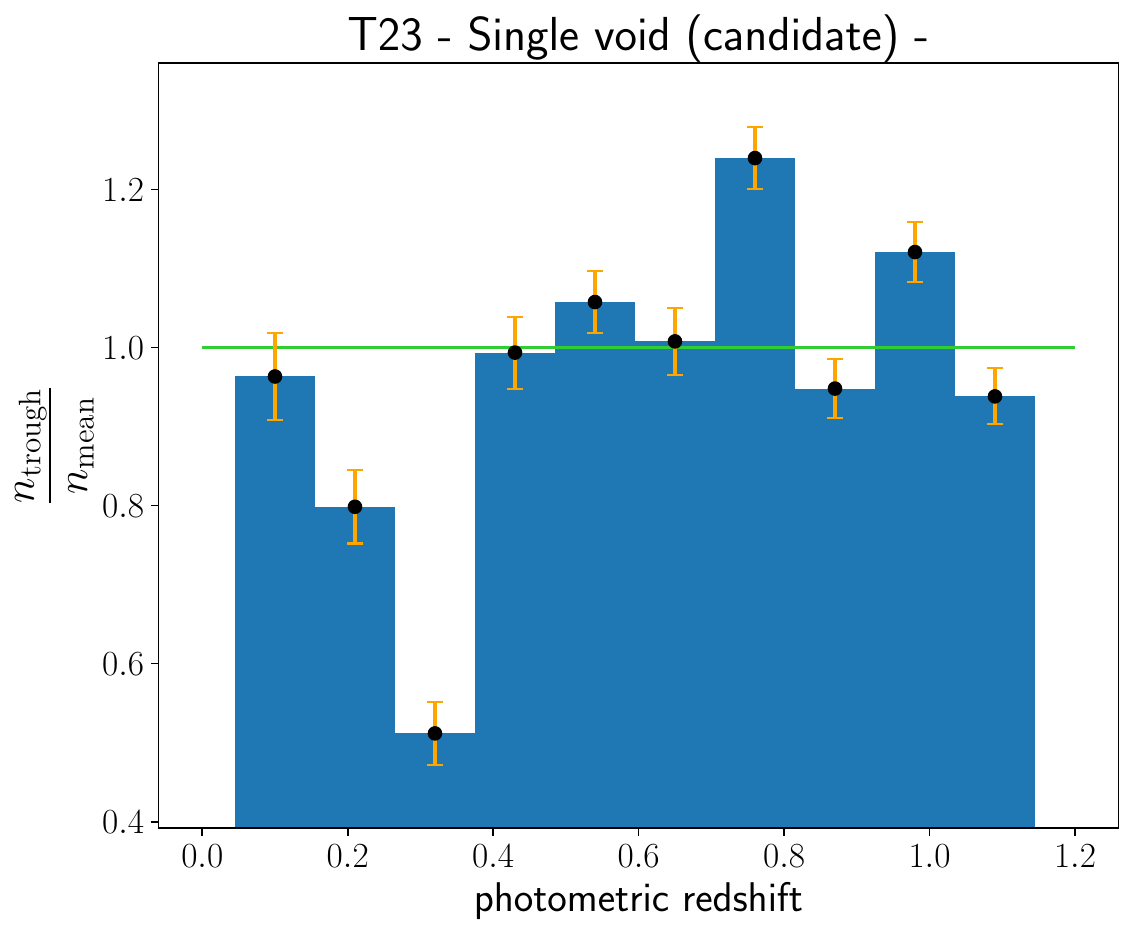}
    \includegraphics[width=0.45\textwidth]{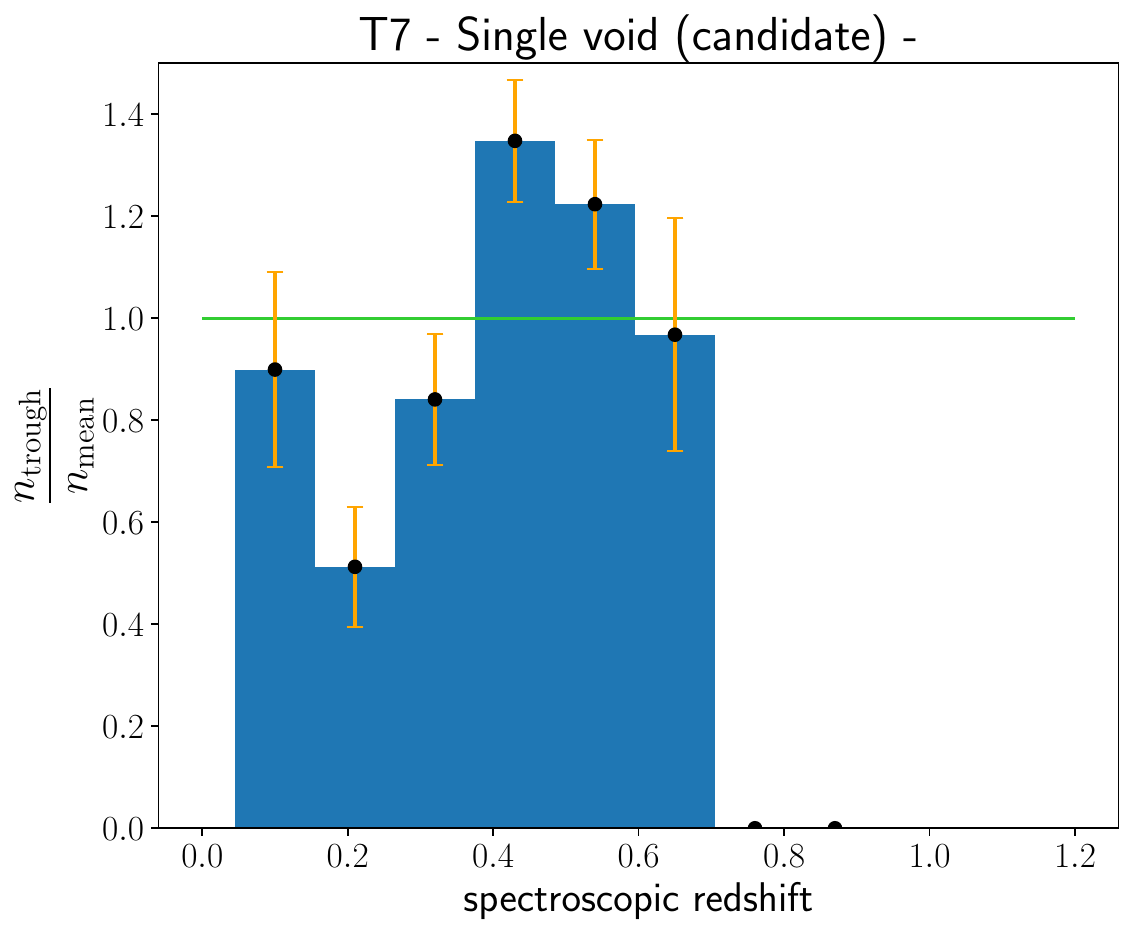}
    \includegraphics[width=0.45\textwidth]{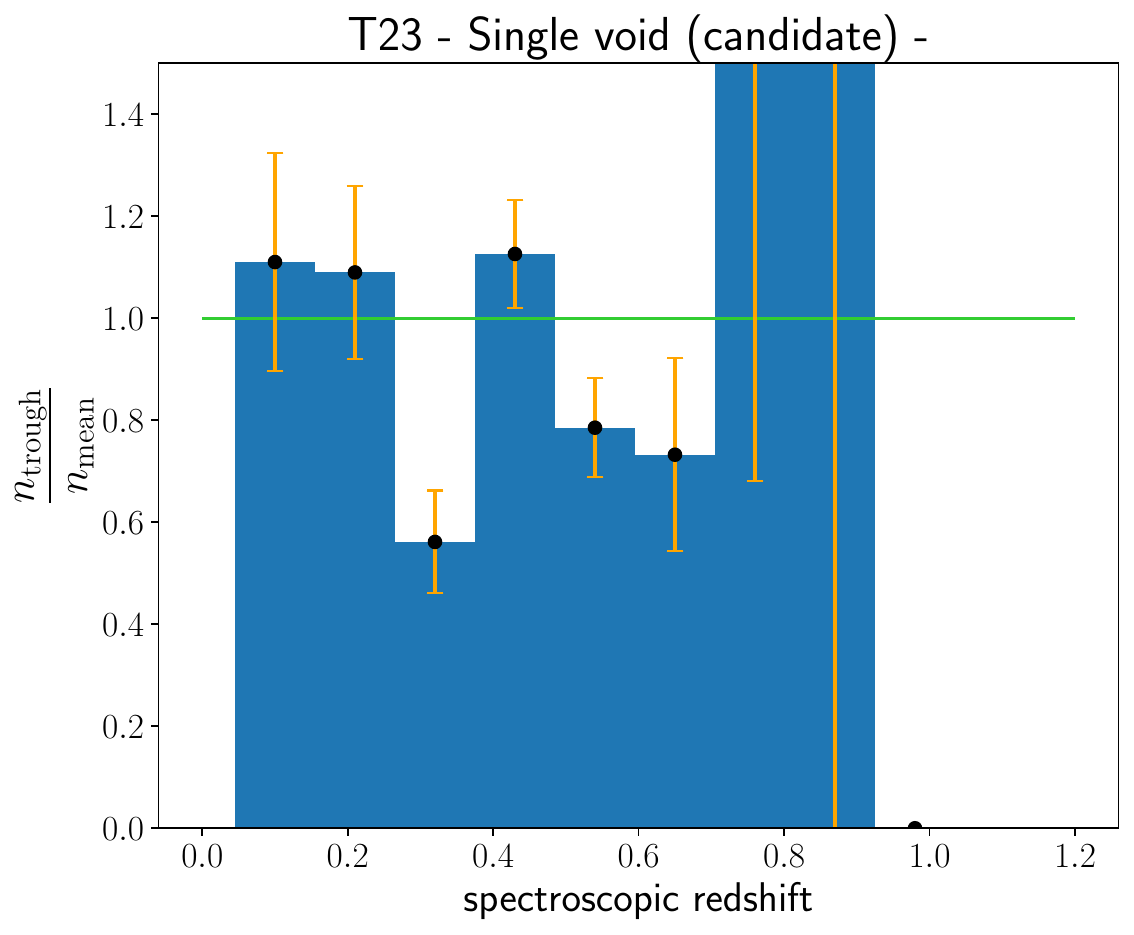}
    \caption{The line-of-sight number density contrasts of HSC photometric LRGs (\textit{upper panels)} and CMASS/LOWZ galaxies (\textit{lower panels}) as a function of redshift for trough T7 (\textit{left panels}) and T23 (\textit{right panels}).
    The error bar corresponds to the Poisson noise from the number of galaxies in each bin.
    The green horizontal line shows the cosmic mean.}
    \label{fig:zdist_single_void}
\end{figure*}

\begin{figure*}
    \includegraphics[width=0.45\textwidth]{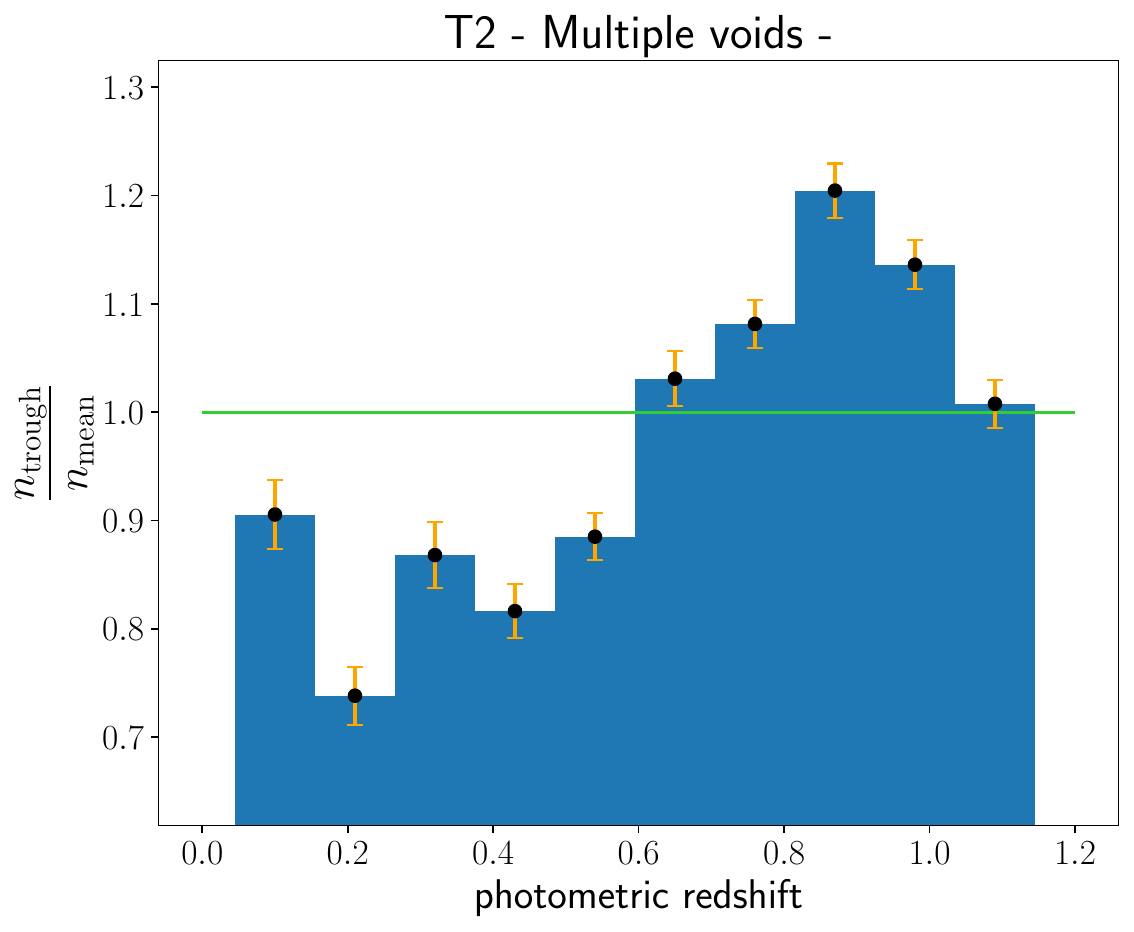}
    \includegraphics[width=0.45\textwidth]{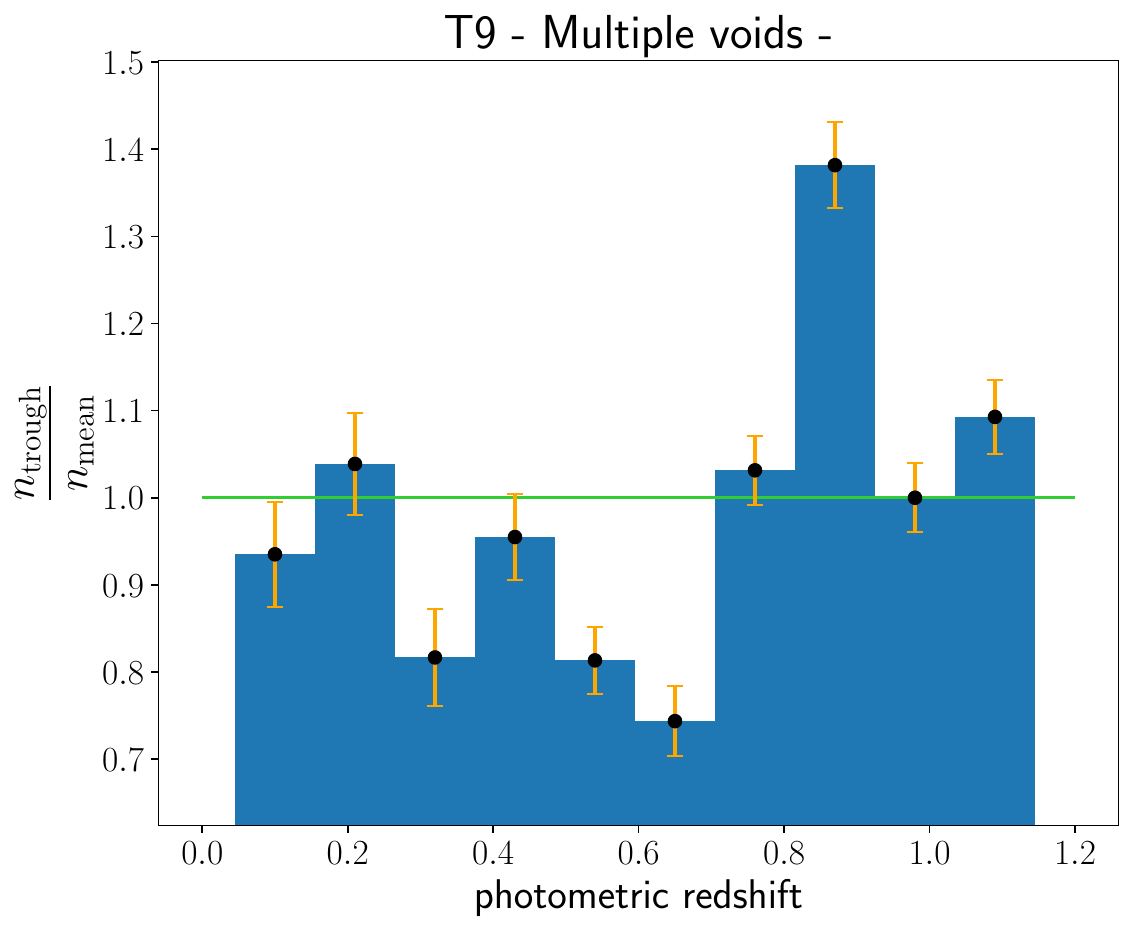}
    \includegraphics[width=0.45\textwidth]{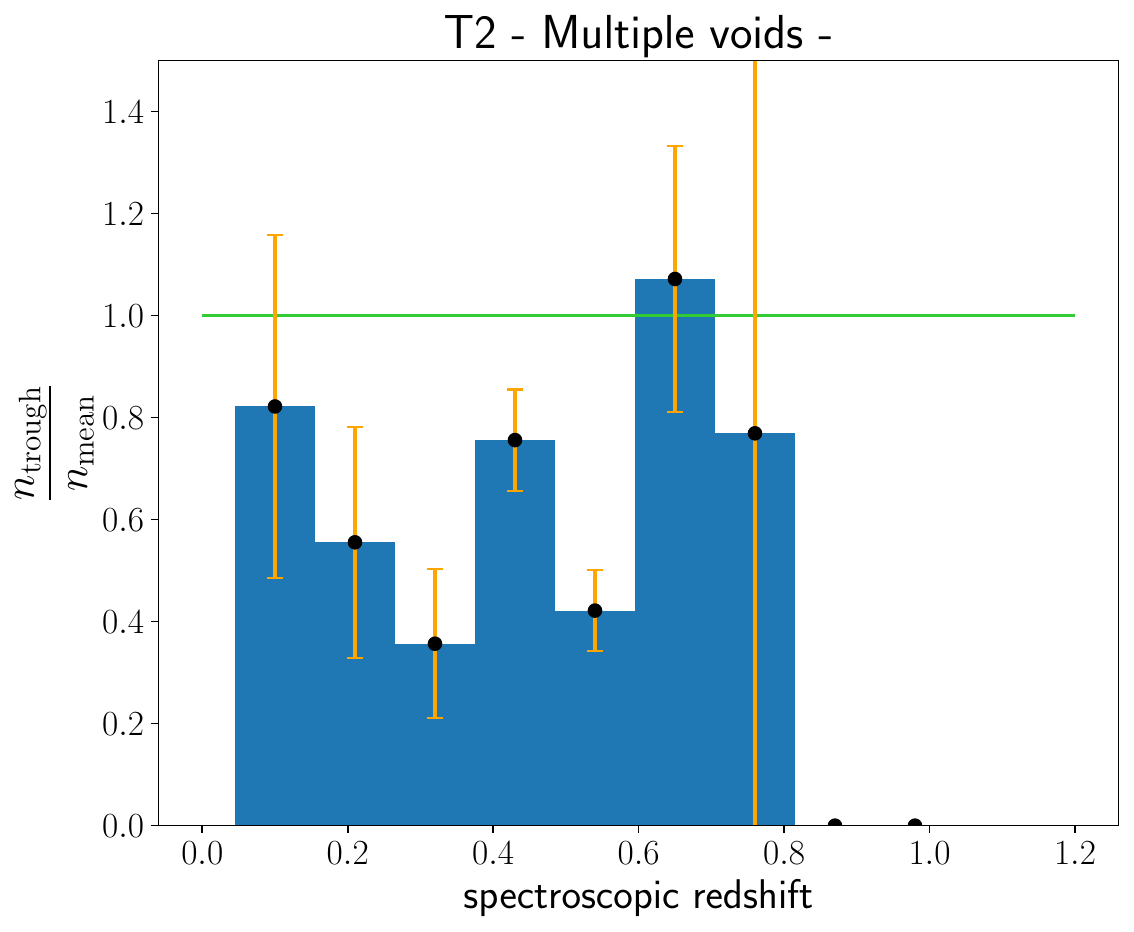}
    \includegraphics[width=0.45\textwidth]{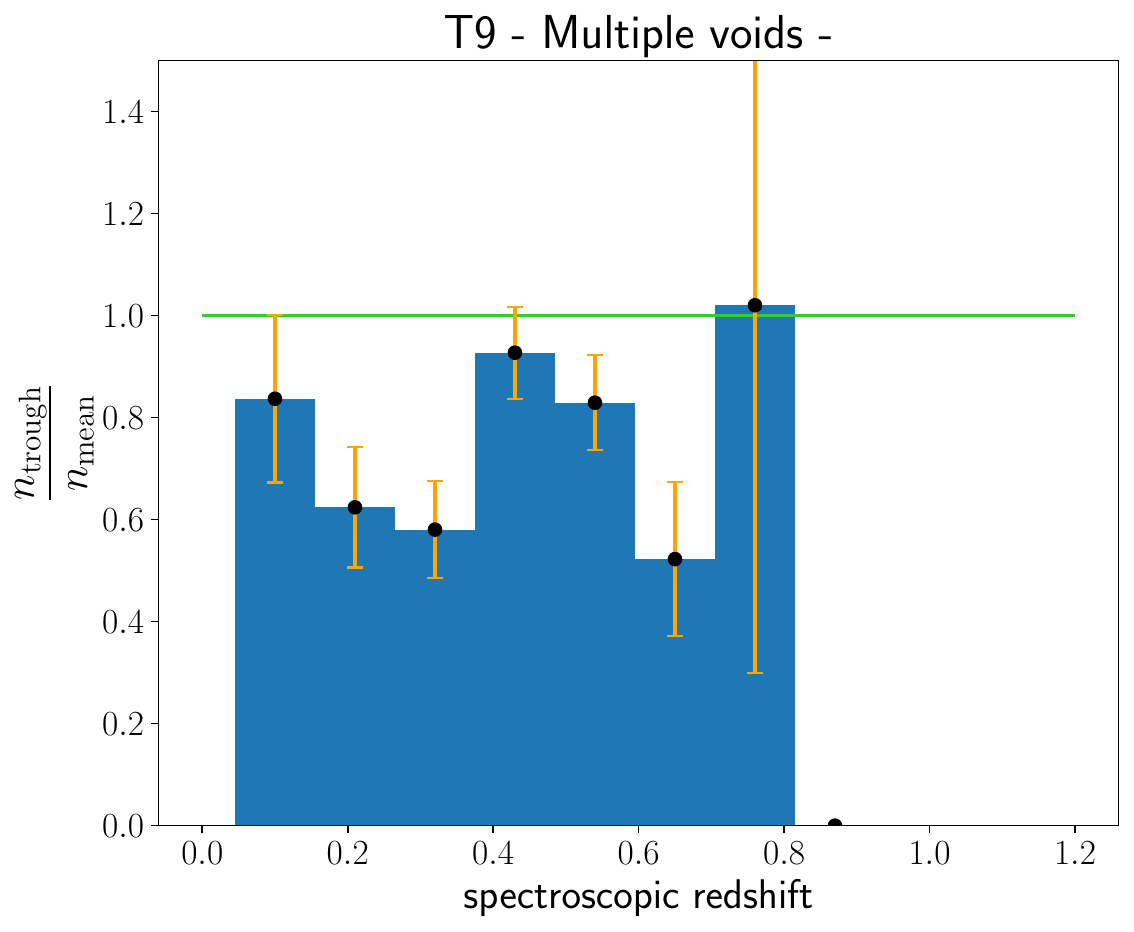}
    \caption{Similar to Figure~\ref{fig:zdist_single_void}, but for T2 (\textit{left panels}) and T9 (\textit{right panels}).}
    \label{fig:zdist_multiple_voids}
\end{figure*}

Figures~\ref{fig:zdist_single_void} and \ref{fig:zdist_multiple_voids}
show the line-of-sight number density contrasts of HSC photometric LRGs
and CMASS/LOWZ galaxies.
A strong depression appears both for LRGs and CMASS/LOWZ
galaxies at $z \simeq 0.2$ for T7 and $z \simeq 0.3$ for T23,
which makes these troughs strong candidates of single-void origin troughs.
In addition, a decrement of the number density for T23 appears at $z \simeq 0.6$ in CMASS/LOWZ. However, the decrement is significant within the $2\sigma$ level, and the decrement of the number density is not observed for HSC galaxies. Therefore, the decrement may be caused by the statistical fluctuation.

On the other hand, both T2 and T9 have multiple decrements at different redshifts in their number density contrasts. Specifically, there are decrements at $z \simeq 0.2$ and $z \simeq 0.5$ for T2 
and at $z \simeq 0.3$ and $z \simeq 0.6$ for T9.
These features indicate that trough signals of T2 and T9 are induced by
multiple voids aligned along the line-of-sight.

\subsection{Three-dimensional mass map around troughs}
An independent and complementary approach to probing the line-of-sight density structure
is the three-dimensional
mass mapping formulated in Section~\ref{sec:3D_mass_mapping}.
This method does not require external data sets such as HSC photometric LRG or CMASS/LOWZ catalogues. Also, it is not affected by the uncertainty of the connection between galaxy and dark matter distributions. However, reconstructed three-dimensional mass maps tend to be noisy, and our reconstruction method described in Section~\ref{sec:3D_mass_mapping} suffers from smearing along the line-of-sight direction as well as the redshift bias in reconstructed mass maps \citep{Simon2009,Oguri2018}. Nevertheless, we study three-dimensional mass maps around single-void origin candidates for further cross-checking.

\begin{figure*}
    \includegraphics[width=\columnwidth]{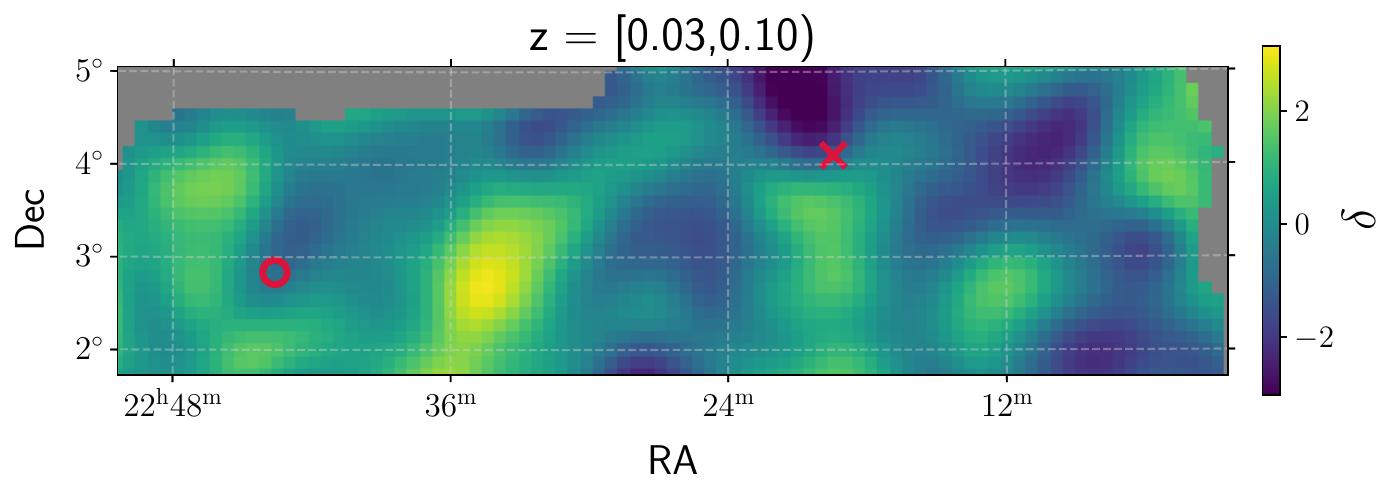}
    \includegraphics[width=\columnwidth]{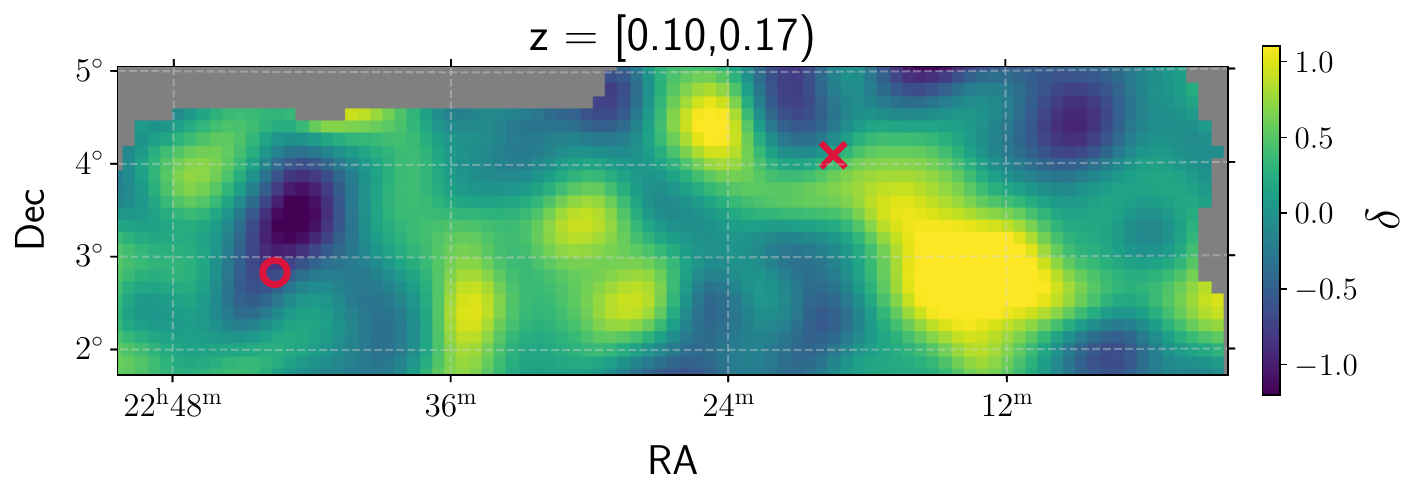}
    \includegraphics[width=\columnwidth]{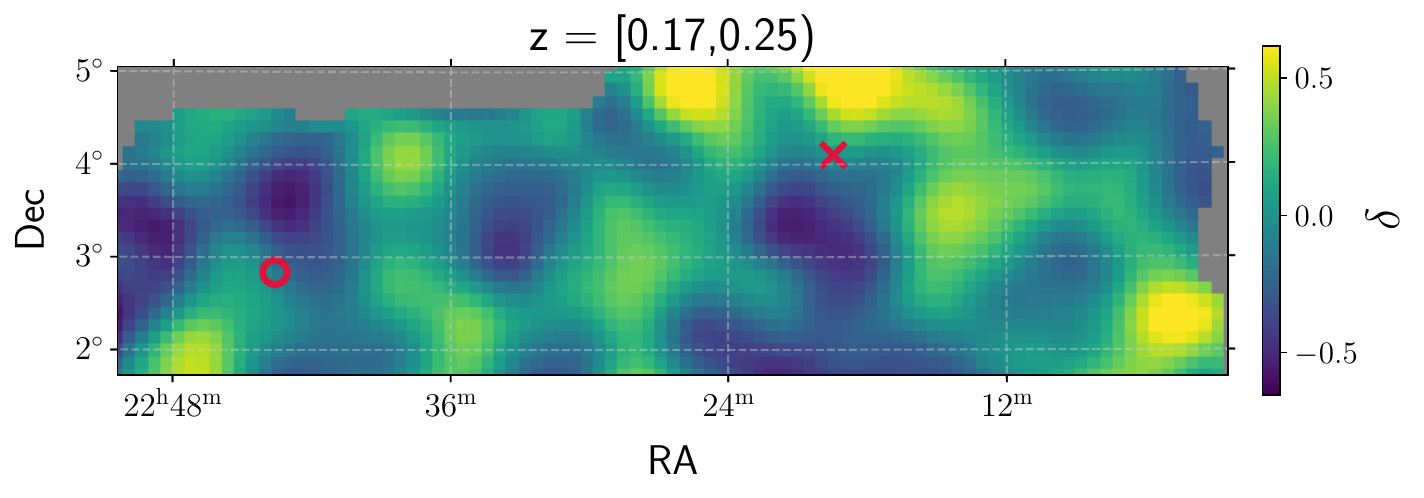}
    \includegraphics[width=\columnwidth]{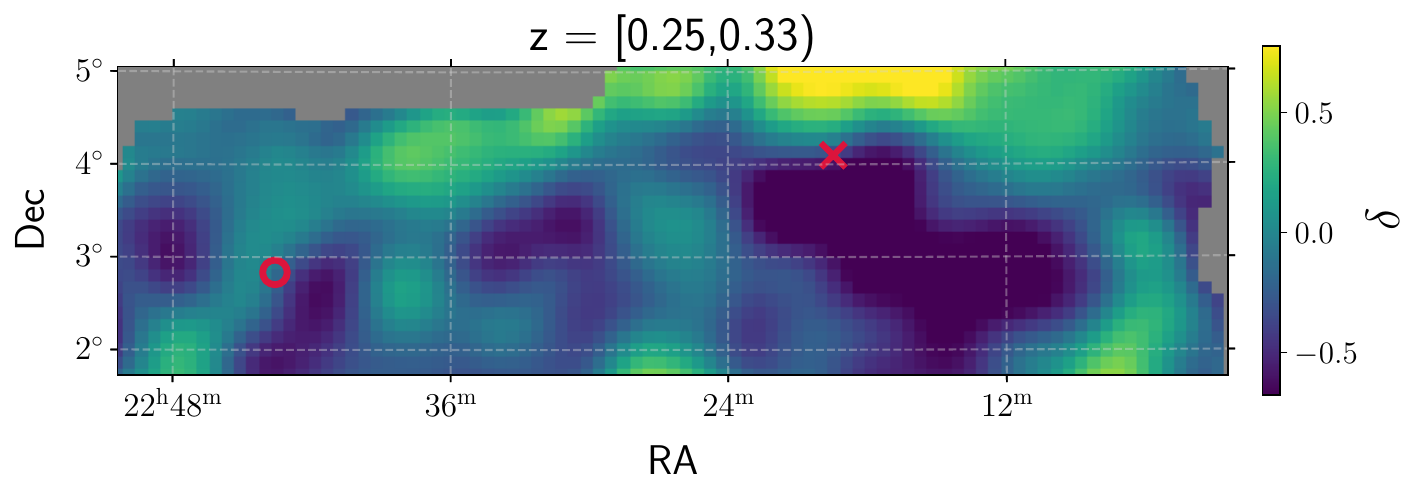}
    \includegraphics[width=\columnwidth]{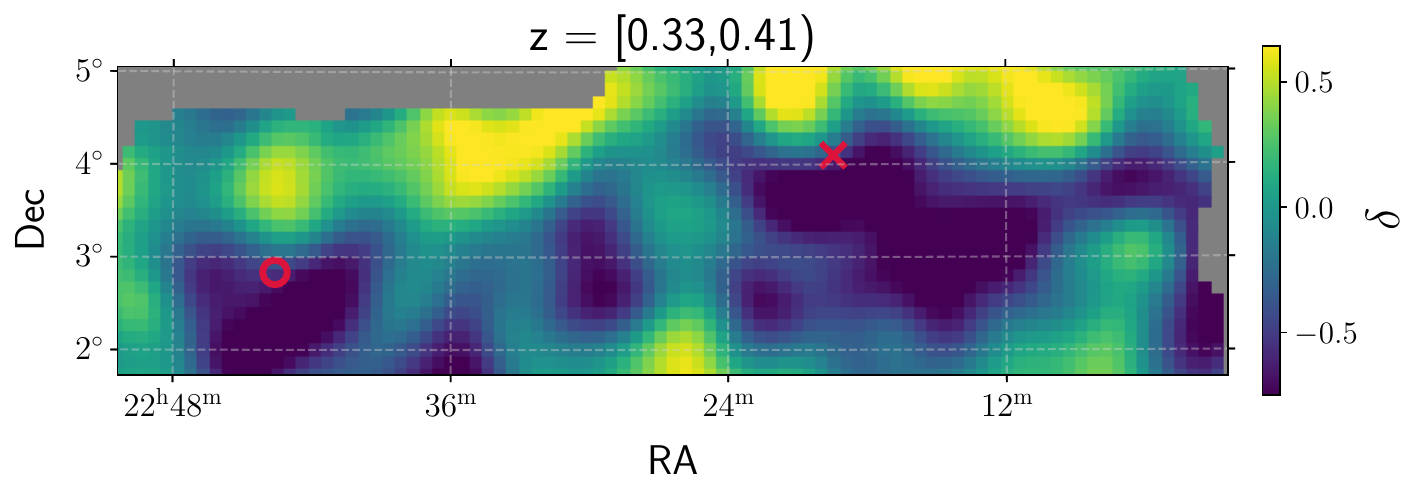}
    \includegraphics[width=\columnwidth]{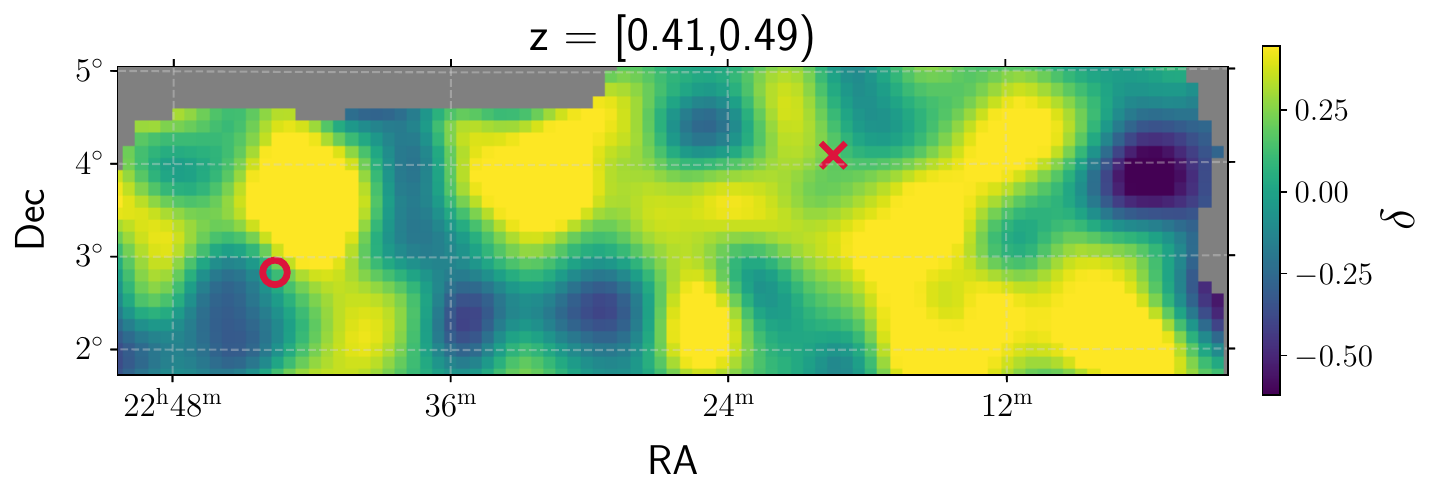}
    \caption{The three-dimensional mass density fluctuation maps covering both of the single-void origin candidates, T7 (\textit{circles}) and T23 (\textit{crosses}), with six redshift bins in the range of $z = 0.03 \text{--} 0.49$.}
    \label{fig:3D_mass_map}
\end{figure*}

Figure~\ref{fig:3D_mass_map} shows the reconstructed three-dimensional mass density fields around the region of single-void origin candidates, T7 and T23.
According to the analysis of line-of-sight number density contrasts,
voids that produce trough signals at T7 and T23 are located at $z \simeq 0.2$
and $z \simeq 0.3$, respectively.
For the trough T23, there is a clear correspondence in the three-dimensional
mass map; the mass density field of $z = 0.25 \text{--} 0.33$
exhibits low-density structures around the trough.
On the other hand, the source void of T7 is ambiguous
from the three-dimensional mass maps.
The density field at $z = 0.17 \text{--} 0.25$ around the trough
is not underdense but rather, low-density regions
are found at $z = 0.10 \text{--} 0.17$ and $z= 0.25 \text{--} 0.33, z = 0.33 \text{--} 0.41$.

\begin{figure*}
    \includegraphics[width=0.45\textwidth]{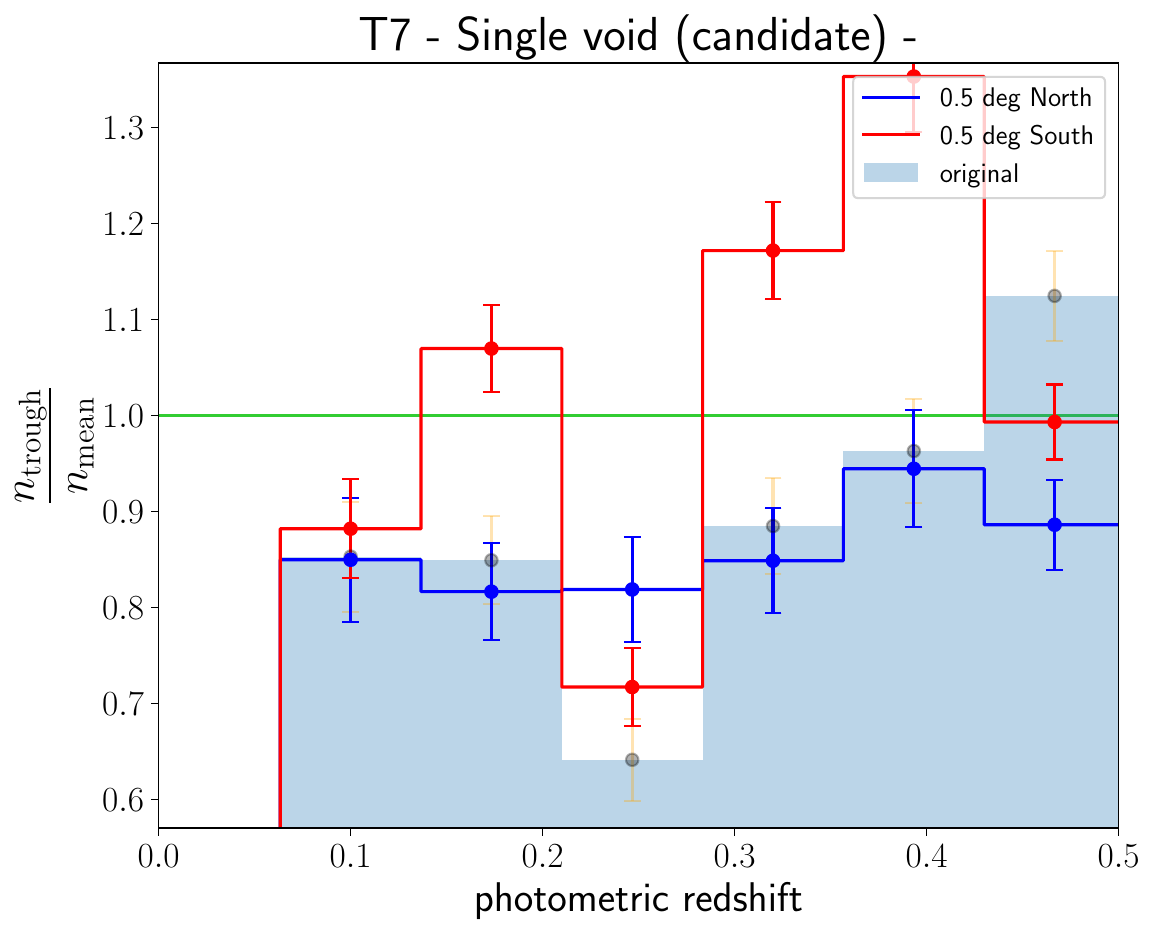}
    \includegraphics[width=0.45\textwidth]{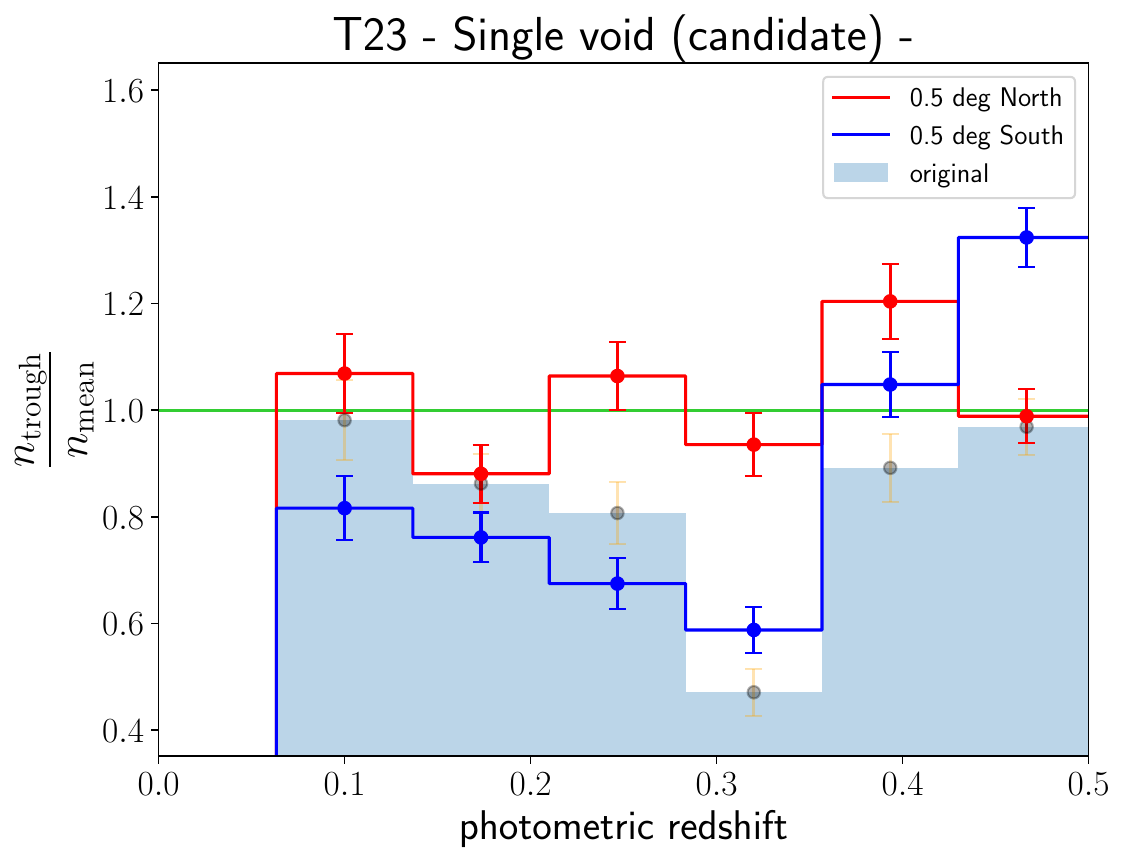}
    \includegraphics[width=0.45\textwidth]{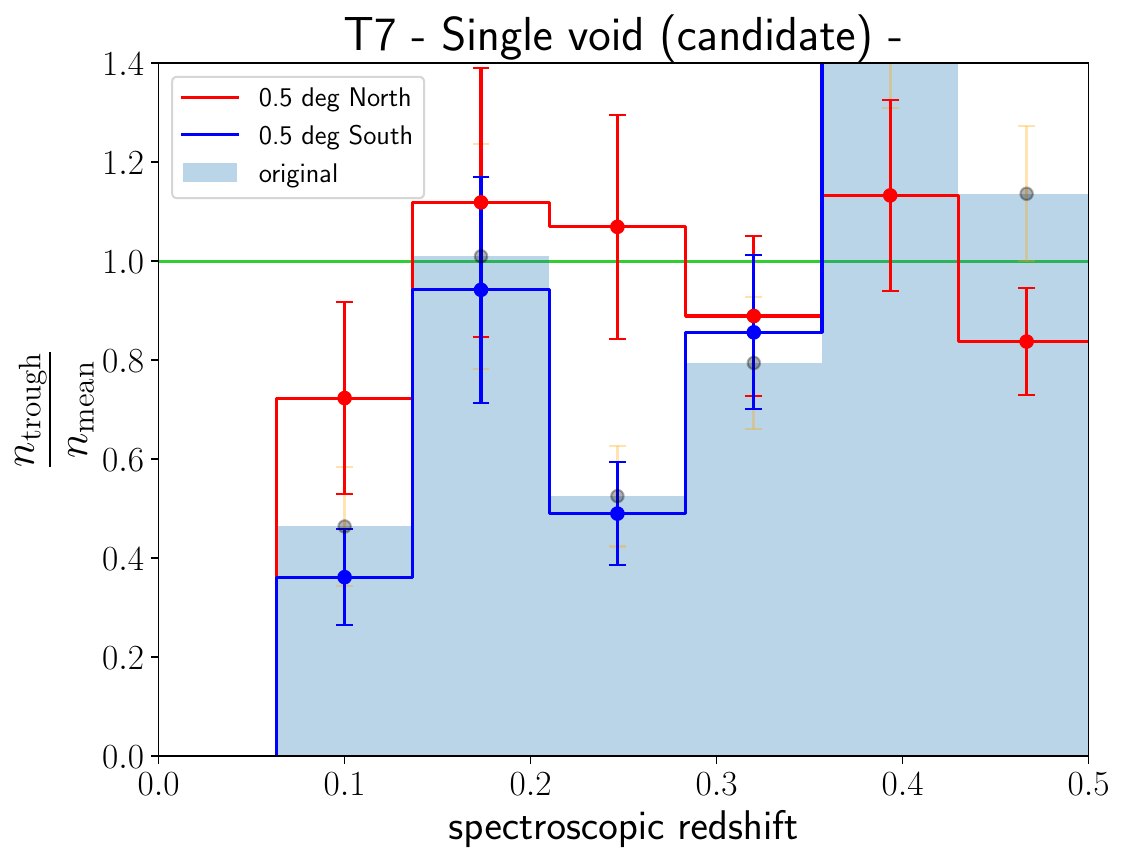}
    \includegraphics[width=0.45\textwidth]{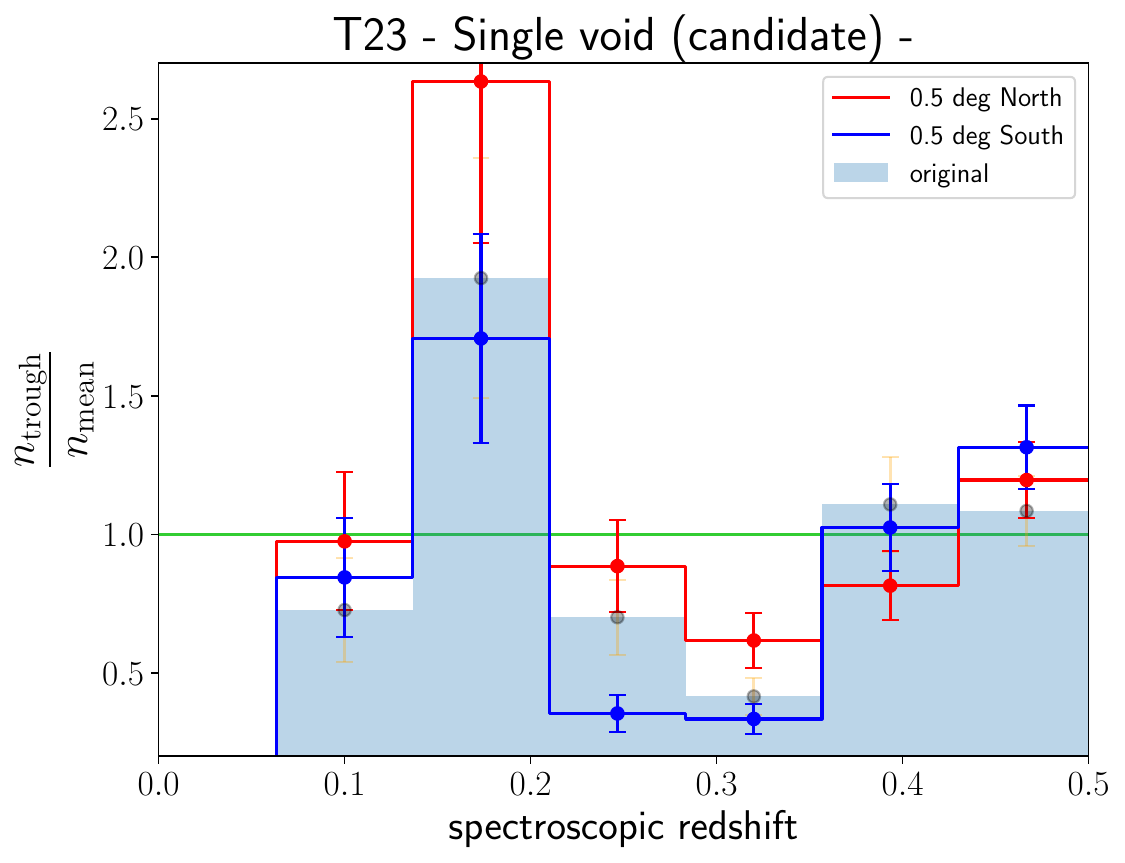}
    \caption{Similar to Figure~\ref{fig:zdist_single_void}, but the redshift distributions shifted from T7 (\textit{left}) and T23 (\textit{right}) to the north (\textit{red}) and south (\textit{blue}) by $0.5\,\mathrm{deg}$ are shown.}
    \label{fig:zdist_single_north_and_south}
\end{figure*}

There is an angular offset between the trough position and the low-density region found
in the three-dimensional mass maps.
The low-density regions appear at $z = 0.10 \text{--} 0.25$ at the northern side from the centre of T7
but also at the southern side at $z = 0.25 \text{--} 0.41$.
A similar feature appears for T23;
the low-density regions are found at $z = 0.03 \text{--} 0.10$
at the northern side and at $z = 0.25 \text{--} 0.41$ at the southern side.
To investigate the global density structure around the troughs,
we measure the redshift distribution of galaxies
at the positions shifted from the centres of T7 and T23
to the north and south by $0.5 \, \mathrm{deg}$ in declination
in Figure~\ref{fig:zdist_single_north_and_south}.
Although three-dimensional mass mapping suffers from large statistical uncertainty,
the low-density region at the south of T23 leads to the decrement of photometric LRGs
and spectroscopic galaxies at redshifts $z \sim 0.2 \text{--} 0.3$.
From the three-dimensional mass mapping, the size of the low-density region is more than $10 \, \hMpc$,
which should be an interesting target for future spectroscopic surveys.

\section{Discussions}
\label{sec:discussions}
Troughs are expected to originate from multiple voids aligned along the line-of-sight because single voids with modest sizes cannot produce significant weak lensing signals \citep{Amendola1999}. Our analysis confirms this expectation by finding that most of the troughs in our sample are multiple-void origin. 

However, our sample of troughs also includes two troughs (T7 and T23) classified as single-void origin candidates. We estimate the weak lensing signals produced by single voids and compare them with observed signals to discuss the possible origin.
For a spherical top-hat void with the radius $r_\mathrm{v}$ and a constant density contrast $\delta_\mathrm{v}$,
the convergence at the void centre $\kappa_\mathrm{v}$ is estimated simply as 
\begin{equation}
    \kappa_\mathrm{v}
    = \frac{2\delta_\mathrm{v} \bar{\rho} (z_\mathrm{v}) r_\mathrm{v}}{\Sigma_{\mathrm{crit}}},
    \label{eq:kappa_void}
\end{equation}
where $\Sigma_\mathrm{crit}$ is the critical surface mass density given by Eq.~\eqref{eq:sigma_crit}.
$\bar{\rho} (z) = (1+z)^3 \Omega_\mathrm{m} \rho_\mathrm{crit}$ is
the mean mass density at the redshift $z$,
$\rho_\mathrm{crit}$ is the critical density in the present Universe,
and $z_\mathrm{v}$ is the redshift of the void.
We take into account the redshift distribution of the source galaxies in the HSC Y3 weak lensing shape catalogue.

Under the assumption of a spherical void, the void radius $r_\mathrm{v}$ can be estimated based on the perpendicular size of the trough in the sky. The radial profile of the convergence around the troughs shown in Figure~\ref{fig:profile_single_void} indicates that the transverse angular size of T7 and T23 is estimated as $1 \, \mathrm{deg}$. Together with the redshift of the void candidates, $z_\mathrm{v}\sim 0.3$, the approximate size of the voids can be estimated as $\sim 11 \, \hMpc$. Since simulations predict that the average central density contrast of voids with this size is quite small, $\delta_\mathrm{v} \sim -1$ \citep{Hamaus2014}, for brevity, we simply assume $\delta_\mathrm{v} = -1$.

Assuming the parameter values mentioned above, from Eq.~\eqref{eq:kappa_void}, we find that the convergence signal at the void centre is computed as $\kappa_\mathrm{v} \sim -1 \times 10^{-3}$.  In contrast, Figure~\ref{fig:profile_single_void} indicates that the observed convergence values at the centres of single-void origin candidates are $\kappa_\mathrm{v} \sim -8 \times 10^{-3}$. To put it another way, the expected S/N from the candidate single void for troughs T7 and T23 is found to be small, with $|\mathrm{S/N}| < 1$, in contrast to the observed S/N of $|\mathrm{S/N}| \sim 6$. This simple analytic estimate confirms the previous argument \citep{Amendola1999} that single voids cannot produce significant signals in weak lensing mass maps.

One of the possible solutions to mitigate the tension is to consider a non-spherical void elongated along the line-of-sight direction.
While the spherical symmetry is assumed in the calculation above, simulations predict that the three-dimensional shape of voids is quite non-spherical \citep[e.g.,][]{Park2007,Platen2008}.
For instance, the central convergence value of a prolate void with axis ratios of $f:1:1$ ($f>1$), when observed along the major axis, is modified as
\begin{equation}
    \kappa_\mathrm{v}
    = \frac{2f\delta_\mathrm{v} \bar{\rho} (z_\mathrm{v}) r_\mathrm{v}}{\Sigma_{\mathrm{crit}}},
    \label{eq:kappa_void_f}
\end{equation}
where $r_\mathrm{v}$ is void size along the minor axis, corresponding to the size of the trough observed on the sky. Considering a highly elongated void with $f\sim 3\,\text{--}\,4$ or an even larger value, the weak lensing signal can be significantly enhanced in proportion to $f$. Although such highly prolate voids are rare, the analysis with $N$-body simulations \citep{Park2007} suggests that there exist a few highly elongated voids with $f\sim 3\,\text{--}\,4$ in the simulation volume of $(500 \,h^{-1} \,\mathrm{Mpc})^3$, which is roughly 10 times smaller than the survey volume of the presented weak lensing analysis. Hence, detecting such prolate voids in HSC weak lensing analysis is statistically consistent with the findings of \citet{Park2007}.

Together with the so-called Eddington bias \citep{Eddington1913}, that can also enhance the observed S/N. However, it is difficult to predict the Eddington bias for voids since the probability distribution function of the trough significance is less known at this moment. Detailed analysis of ray-tracing simulations is needed to quantitatively address the Eddington bias for troughs, which is beyond the scope of this paper. On the other hand, for shear-selected clusters identified from high S/N peaks in mass maps, the Eddington bias of the cluster masses is shown to be up to around 55\%, which were derived based on detailed simulations \citep{Chen2020}. A similar level of the Eddington bias may be expected for troughs.

We conclude that weak lensing signals of troughs T7 and T23 can be explained by single voids elongated along the line-of-sight. This hypothesis can be tested with dense spectroscopic follow-up observations in the candidate trough regions with, e.g., Prime Focus Spectrograph \citep{Takada2014}.
We present the troughs with high $|\mathrm{S/N}| > 7$ and medium $ |\mathrm{S/N}|>5.7$ significance levels. Obviously, a low threshold will yield more troughs. For example, we have identified 300 troughs with $|\mathrm{S/N}| > 2$. However, the 3D density structures and the redshift of the voids are determined with a visual inspection of HSC and SDSS galaxy number densities, which hampers thorough analysis of less significant voids. We leave the systematic analysis of the full sample of troughs for future work.

The selection effect can easily explain the alignment of elongated voids along the line-of-sight. Since the weak lensing signal is significantly enhanced when observed along the major axis, the major axes of single voids found from mass maps should be preferentially aligned along the line-of-sight direction. This is indeed the case for peaks selected in mass maps. \citet{Hamana2012} confirm the presence of an orientation bias for weak lensing selected clusters through a meticulous analysis of ray-tracing simulations. They also find that this bias is stronger for higher S/N peaks. Since our trough sample represents the most significant troughs with the highest $|\mathrm{S/N}|$, the orientation bias is expected to be quite strong.

\section{Conclusions}
\label{sec:conclusions}
Voids are large-scale underdense regions in the Universe and
one of the building blocks of the cosmic web structure.
Since there are few or no galaxies in voids,
the void statistics is expected to be less affected by baryonic physics and hence offers a powerful approach
to constrain the geometry of the Universe through Alcock--Paczynski test
\citep{Lavaux2012,Hamaus2016,Mao2017,Nadathur2019,Hamaus2020}
and to measure the BAO scale
\citep{Kitaura2016,Liang2016,Nadathur2019,Zhao2020,Zhao2022}.
However, the scarcity of galaxies in voids
makes it challenging to identify voids using spectroscopic galaxy surveys.
Weak gravitational lensing has the potential to overcome this fundamental problem
because it probes the density field in an unbiased manner.

On the other hand, searching for voids with weak lensing has been considered problematic because single voids typically cannot produce significant weak lensing signals.
In this paper, we have performed the mass mapping analysis with the latest
Subaru HSC Y3 shape catalogue \citep{Li2022} with the survey area of $433.48 \, \mathrm{deg}^2$ and the mean source galaxy number density of $22.9 \, \mathrm{arcmin}^{-2}$ to identify troughs,
which are defined as local minima in weak lensing mass maps and are created by aligned underdense regions, i.e. voids.
Excluding troughs near the edge of the survey footprint, we have identified 4 high-significance troughs with $|\mathrm{S}/\mathrm{N}| > 7$ 
and 11 medium-significance troughs with $|\mathrm{S}/\mathrm{N}| > 5.7$. 
To study the line-of-sight structure of these troughs, 
we have utilized redshift distributions of two galaxy samples, HSC photometric LRGs \citep{Oguri2014,Oguri2018a,Oguri2018}
and spectroscopic CMASS/LOWZ galaxies from SDSS DR12 \citep{Reid2016}.
We have investigated the line-of-sight density structures at the trough positions
by cross-checking both galaxy samples.
While multiple underdense regions are seen in the redshift
distributions of the galaxies for most of the troughs, we find that there are two troughs for which only single underdense regions are identified in their redshift distributions of the galaxies albeit relatively high threshold $|\mathrm{S/N}| > 5.7$.
Weak lensing signals of these two troughs are potentially produced by single voids.
We have also carried out the three-dimensional mass mapping \citep{Simon2009,Oguri2018}
around these two troughs to investigate the global density structures.
From this analysis, we have identified large-scale ($\gtrsim 10 \, \hMpc$)
underdense regions around one of the troughs.

The convergence radial profiles of these two troughs indicate that their perpendicular physical sizes are $\sim 11 \, \hMpc$. Assuming a zero-density spherical void, this translates into the predicted convergence at the void centre of $\sim -1 \times 10^{-3}$, which is much larger than the observed convergence at the centres of the troughs, $\sim -8 \times 10^{-3}$. 
We have argued that these two troughs can still originate from single voids if we consider single voids highly elongated along the line-of-sight direction.
Simulations predict that voids are not spherical but rather prolate, and their weak lensing signals can be significantly boosted when their major axes are aligned with the line-of-sight direction. In addition, the Eddington bias \citep{Eddington1913} can also help mitigate the difference between the predicted and observed convergence values. 
Our hypothesis can be tested with dense spectroscopic follow-up observations in the trough regions to identify voids with great accuracy, which we leave for future work.

\section*{Acknowledgements}
We thank an anonymous referee for useful comments.
This work was supported in part by MEXT/JSPS KAKENHI Grant Number
JP22K14036 (K.O.), JP20H05856 (M.O.), JP23H00108 (A.N.), and JP22K21349 (M.O. and A.N.).

Some of the results in this paper have been derived using the \texttt{healpy} and \texttt{HEALPix} packages.

The Hyper Suprime-Cam (HSC) collaboration includes the astronomical communities of Japan and Taiwan, and Princeton University.  The HSC instrumentation and software were developed by the National Astronomical Observatory of Japan (NAOJ), the Kavli Institute for the Physics and Mathematics of the Universe (Kavli IPMU), the University of Tokyo, the High Energy Accelerator Research Organization (KEK), the Academia Sinica Institute for Astronomy and Astrophysics in Taiwan (ASIAA), and Princeton University.  Funding was contributed by the FIRST program from the Japanese Cabinet Office, the Ministry of Education, Culture, Sports, Science and Technology (MEXT), the Japan Society for the Promotion of Science (JSPS), Japan Science and Technology Agency  (JST), the Toray Science  Foundation, NAOJ, Kavli IPMU, KEK, ASIAA, and Princeton University.
 
This paper is based on data collected at the Subaru Telescope and retrieved from the HSC data archive system, which is operated by Subaru Telescope and Astronomy Data Center (ADC) at NAOJ. Data analysis was in part carried out with the cooperation of Center for Computational Astrophysics (CfCA) at NAOJ.  We are honored and grateful for the opportunity of observing the Universe from Maunakea, which has the cultural, historical and natural significance in Hawaii.
 
This paper makes use of software developed for Vera C. Rubin Observatory. We thank the Rubin Observatory for making their code available as free software at http://pipelines.lsst.io/. 
 
The Pan-STARRS1 Surveys (PS1) and the PS1 public science archive have been made possible through contributions by the Institute for Astronomy, the University of Hawaii, the Pan-STARRS Project Office, the Max Planck Society and its participating institutes, the Max Planck Institute for Astronomy, Heidelberg, and the Max Planck Institute for Extraterrestrial Physics, Garching, The Johns Hopkins University, Durham University, the University of Edinburgh, the Queen’s University Belfast, the Harvard-Smithsonian Center for Astrophysics, the Las Cumbres Observatory Global Telescope Network Incorporated, the National Central University of Taiwan, the Space Telescope Science Institute, the National Aeronautics and Space Administration under grant No. NNX08AR22G issued through the Planetary Science Division of the NASA Science Mission Directorate, the National Science Foundation grant No. AST-1238877, the University of Maryland, Eotvos Lorand University (ELTE), the Los Alamos National Laboratory, and the Gordon and Betty Moore Foundation.

\section*{Data Availability}
The catalogue and data underlying this article are available on the
public data release site of Hyper Suprime-Cam Subaru Strategic Program
(\url{https://hsc.mtk.nao.ac.jp/ssp/data-release}).
The HSC Y3 shape catalogue will be available
at the same site later.



\bibliographystyle{mnras}
\bibliography{main}





\bsp	
\label{lastpage}
\end{document}